\documentclass[12pt]{iopart}
\usepackage{iopams}  
\usepackage{lipsum}
\usepackage{graphicx}

\begin{document}

\title{Comparison between Jacobi-Anger and saddle point methods to treat Above-threshold ionization}
\author{Danish Furekh Dar and Stephan Fritzsche}

\address{Helmholtz-Institut Jena, Fr\"o{}belstieg 3, D-07743 Jena, Germany,\\
GSI Helmholtzzentrum f\"ur Schwerionenforschung GmbH, Planckstrasse 1, D-64291 Darmstadt, Germany,\\
Theoretisch-Physikalisches Institut, Friedrich-Schiller-Universit\"at Jena, Max-Wien-Platz 1, D-07743 Jena, Germany.}
\ead{danish.dar@uni-jena.de}
\vspace{10pt}
\begin{indented}
\item[]November 2024 
\end{indented}

\begin{abstract}
	We present a detailed comparison of theoretical approaches for modeling strong-field ionization by few-cycle laser pulses. The dipole approximation is shown to accurately capture interference structures in photoelectron spectra, while non-dipole effects introduce significant momentum shifts along the propagation direction. Two complementary analytical methods are used: the Jacobi-Anger expansion provides complete spectral decomposition of transition amplitudes, whereas the saddle-point method efficiently identifies dominant ionization pathways. Through this comparative study within the strong-field approximation framework, we establish validity conditions and practical advantages for each approach. Our results provide guidelines for selecting theoretical methods for advancing the interpretation of strong-field processes. These findings provide a roadmap for interpreting strong-field ionization spectra and momentum distributions, highlighting where non-dipole effects and method choice critically alter predictions.
\end{abstract}

%
\maketitle
%


%

\section{Introduction}
When an atom is subjected to an intense laser field, it can absorb multiple photons, leading to ionization. This process often extends beyond the minimal number of photons required for ionization, resulting in the absorption of additional photons. This phenomenon, known as Above-Threshold Ionization (ATI), was first experimentally observed by Agostini and his co-workers in 1979 \cite{Agostini1979}. ATI is typically studied through the analysis of photoelectron kinetic energy distributions, often employing advanced techniques such as time-of-flight spectroscopy \cite{Paulus2001,Paulus2003}. With advancements in laser technology, including higher intensities, shorter pulse durations, and improved detection methods, the complexity and resolution of ATI spectra have significantly increased. These developments have introduced new challenges in fully understanding the underlying physical mechanisms \cite{Treiber2022}.

Typically, an ATI spectra is characterized by a series of equally spaced peaks, each separated by the photon energy, with intensities decreasing for higher-order peaks \cite{JHEberly_1988,Bashkansky1988}. This pattern can be well-explained by perturbation theory, although the probability of absorbing additional photons gradually decreases with each successive photon \cite{Bachau2013,Tilley_2015,Varma2009}. However, the advent of more powerful lasers and ultrafast pulses revealed more intricate spectral features. Phenomena such as channel closing \cite{Cormier_2001}, where certain ionization pathways become energetically inaccessible, and Freeman resonances \cite{Freeman1987}, arising from intermediate resonant states, began to distort the lower-order peaks and introduced fine structures within the spectra. These observations underscored the dynamic interplay between the atom's energy levels, modified by laser-induced ac Stark shifts, and the motion of free or nearly free electrons within the oscillating laser field.

A key concept in understanding these dynamics is the ponderomotive energy ($U_p$), which represents the average kinetic energy of a free electron oscillating in an electromagnetic field. The ponderomotive energy significantly influences the ionization dynamics and the final photoelectron momentum distributions, particularly in high-intensity regimes \cite{Bucksbaum1987,Agostini1987}.The energy spacing and structure of ATI peaks, contributing to more complex spectral features. Modern research continues to explore these intricate interactions using advanced computational models and experimental setups, aiming to deepen our understanding of strong-field ionization processes \cite{Kling2018,Ooi2017,Picca_2015,Rottke1994}.

ATI phenomena are typically categorized into two distinct ionization regimes: multiphoton ionization and tunneling ionization. These regimes are distinguished by the Keldysh parameter \cite{Keldysh1964}, $\gamma$, defined as \(\gamma = \sqrt{\frac{I_p}{2 U_p}},\) where $I_p$ is the ionization potential and $U_p$ is the ponderomotive energy, which depends on the peak laser field strength $E_0$ and the laser frequency $\omega$. When $\gamma \gg 1$, ionization of atoms predominantly occurs via the multiphoton process, resulting in an ATI spectrum characterized by evenly spaced narrow peaks, each separated by the photon energy \cite{Agostini1979}. This structure reflects the quantized nature of photon absorption and can be interpreted as arising from the interference of electrons ionized during different cycles of the laser pulse \cite{Arbó_2007}.

To explain strong-field ionization phenomena, various theoretical models have been developed. Beyond the direct numerical solutions of the time-dependent Schrödinger equation (TDSE) and semiclassical electron dynamics models \cite{torlina2012,javanainen1988,jheng2022,bauer2006,madsen2007,patchkovskii2016}, the strong-field approximation (SFA) has emerged as a robust method for predicting ATI and high-harmonic generation (HHG) spectra under diverse laser conditions \cite{Keldysh1964,Faisal_1973,Reiss1980}. The SFA simplifies the problem by neglecting the Coulomb potential of the parent ion once the electron enters the continuum, assuming this potential is insignificant compared to the external laser field. Additionally, the model assumes the laser field has no impact on the atom's initial state, treating it as a stationary state unaffected by external influences. These simplifications facilitate analytical calculations of photoelectron momentum distributions resulting from ionization.

Modern laser laboratories now possess intense laser pulses that last only a few optical cycles \cite{Bucksbaum2003,Krausz}. These ultrashort pulses serve as powerful tools for probing atomic and molecular dynamics over extremely brief timescales. For example, combining few-cycle laser pulses with subfemtosecond soft X-ray pulses has enabled the observation of atomic processes on the attosecond scale \cite{Uiberacker2007}. Furthermore, high-harmonic generation with few-cycle pulses has been instrumental in studying nuclear motion within molecules and imaging electronic orbitals \cite{LaraAstiaso2018}.

The quantum mechanical transition amplitudes for ATI and other strong-field phenomena are expressed as multidimensional integrals involving highly oscillatory functions. These complex integrals can be evaluated using numerical techniques or asymptotic methods. The asymptotic approach not only reduces computational complexity but also provides insights into the distinct contributions of various quantum pathways, each associated with different solutions of the saddle-point equations \cite{Jašarević_2020,NAYAK20191}. Over the past three decades, the saddle-point method has been extensively applied to compute transition amplitudes for a variety of strong-field processes \cite{Habibovi2023,Milošević2016,Weber2021,Minneker2021}.

In recent years, numerous experimental investigations have focused on observing and modeling nondipole effects in strong-field ionization, particularly at near-infrared \cite{smeenk2011, hartung2019, Haram2019, Lin2022} and mid-infrared wavelengths \cite{ludwig2014, maurer2018, Willenberg2019}. These nondipole effects are often manifested as a shift in the peak of the photoelectron momentum distribution (PMD) along the laser's propagation direction, deviating from the zero-momentum peak predicted by the dipole approximation. Theoretical studies addressing these effects have employed approaches such as the SFA, the TDSE, and tunneling models \cite{Dar2023, dar-atoms, Fritzsche2022, Boning2022, Boning2019, Minneker2022, Jensen2020, Ni2020, Brennecke2021, Lund2021, He2022, Klaiber2022, Madsen2022, Liang2022, Kahvedzic2022}. The inadequacy of the electric dipole approximation in this context has been well established and is primarily attributed to radiation pressure and magnetic field interactions \cite{Reiss2008,Reiss2013}.

In this paper, we conduct a detailed study of ATI by employing two methods: the Jacobi-Anger expansion and the saddle-point method. We provide a comprehensive analysis of the ionization spectrum and momentum distributions resulting from these two approaches, highlighting their respective advantages and limitations. Additionally, we explore the nondipole effects arising from the spatial structure of the laser pulse, offering insights into how these effects influence the overall ionization dynamics. For the sake of simplicity, we focus on the hydrogen-like 1\textit{s} state while performing calculations for the argon atom. This choice allows us to isolate and examine the fundamental processes without the added complexity of multielectron interactions. By systematically evaluating these approaches, we establish their respective validity regimes and discuss their advantages and limitations. This study advances the understanding of ionization dynamics beyond the dipole approximation, providing insights for refining theoretical models in strong-field physics. Our findings have implications for precision measurements and ultrafast laser-matter interactions, offering guidance for future studies in attosecond science.


\section{Theory}
In a standard ATI experiment, a gaseous sample of atoms is subjected to an intense laser field, resulting in the emission of electrons. The key observable in such experiments is the angle-resolved photoelectron momentum distribution (PMD), which quantifies how electrons are ejected in different directions with varying momenta. This distribution is mathematically described by the energy- and angle-differential photoionization probability \( \mathbb{P}(\mathbf{p}) \), where \( \mathbf{p} \) denotes the momentum of the emitted photoelectron detected.

The primary objective of a theoretical framework for ATI is to accurately calculate \( \mathbb{P}(\mathbf{p}) \) for a given atomic system interacting with a specific laser field. This process begins by solving the time-dependent Schrödinger equation, which governs the quantum evolution of the electron's wavefunction \( |\psi(t)\rangle \). The Hamiltonian operator \( \hat{H} \) in this context incorporates the kinetic energy of the electron, its interaction with the external laser field, and the potential energy due to the atomic core. The Hamiltonian is given by

\begin{equation}
	\hat{H} = \frac{1}{2} \left( \hat{\mathbf{p}} - q\mathbf{A}(\mathbf{r}, t) \right)^2 + q\phi(\mathbf{r}, t) + V(\mathbf{r}),
	\label{EQ:Hamiltionain}
\end{equation}

where \( q = -e \) is the electron's charge, \( \phi(\mathbf{r}, t) \) and \( \mathbf{A}(\mathbf{r}, t) \) represent the scalar and vector potentials of the laser field, respectively, and \( V(\mathbf{r}) \) is the atomic binding potential. This equation captures the motion of the electron and how it is affected by the oscillating electric and magnetic fields of the laser.

To solve this time-dependent equation, an initial wavefunction \( |\psi(t_0)\rangle \) must be specified, typically representing the bound state of the electron just before the interaction begins. As the system evolves under the influence of the laser field, the probability of detecting an electron with a specific momentum \( \mathbf{p} \) at the detector is determined by projecting the final state \( |\psi(t \to \infty)\rangle \) onto the continuum state \( |\psi_{\mathbf{p}}(t)\rangle \), which corresponds to a free electron with momentum \( \mathbf{p} \).

The differential photoionization probability, which describes the likelihood of emitting an electron with energy \( \epsilon_p = \frac{\mathbf{p}^2}{2} \) into a solid angle element \( d\Omega_{\mathbf{p}} \), is expressed as
\begin{equation}
\mathbb{P}(\mathbf{p}) = |T(\mathbf{p})|^2 \frac{d^3\mathbf{p}}{d\Omega_{\mathbf{p}} d\epsilon_p} = p |T(\mathbf{p})|^2,
\label{EQ:differentialphotoionizationProbability}
\end{equation}
where \( T(\mathbf{p}) \) is the transition amplitude of the process, quantifying the strength of the transition from the bound state to the continuum state. The transition amplitude is defined by
\begin{equation}
	T(\mathbf{p}) = \lim_{t \to \infty} \langle \psi_{\mathbf{p}}(t) | \psi(t) \rangle.
	\label{EQ:trasnisitionamplitude}
\end{equation}
This formalism emphasizes how the interaction between the electron and the intense laser field drives the ionization process. The vector potential \( \mathbf{A}(\mathbf{r}, t) \) and scalar potential \( \phi(\mathbf{r}, t) \) play critical roles in shaping the electron's trajectory and final momentum distribution. At high laser intensities, non-perturbative effects emerge and necessitate advanced theoretical tools like the SFA and saddle-point methods for accurate modeling of ionization dynamics.


\subsection{Transition amplitude}
To understand the transition amplitude, one must analyze the evolution of an electron's wave function under an external field. The wave function $|\psi(t)\rangle$ evolves according to the Hamiltonian $\hat{H}$, where the unitary time-evolution operator $\hat{U}(t, t_0)$ determines the transition from an initial state $|\psi(t_0)\rangle$ to the final state
\begin{equation}\label{eq:wave_function}
	|\psi(t)\rangle = \hat{U}(t, t_0)|\psi(t_0)\rangle.
\end{equation}
The evolution operator $\hat{U}(t, t_0)$ satisfies the well-known time-dependent Schr\"odinger equation, which describes the fundamental dynamics of quantum systems
\begin{equation}\label{eq:schrodinger}
	i \frac{\partial}{\partial t} \hat{U}(t, t_0) = \hat{H}(t)\hat{U}(t, t_0).
\end{equation}
The evolution operator adheres to the normalization condition $\hat{U}(t_0, t_0) = 1$ and satisfies the composition principle, which enables its factorization at intermediate times
\begin{equation}\label{eq:composition}
    \hat{U}(t, t_0) = \hat{U}(t, \tau) \hat{U}(\tau, t_0), \quad \forall \tau.
\end{equation}
When the total Hamiltonian decomposes into two distinct components
\begin{equation}\label{eq:hamiltonian_decomp}
    \hat{H} = \hat{H}_1 + \hat{H}_2,
\end{equation}
the Dyson series expansion provides a means to express the full evolution operator in terms of contributions from these components
\begin{equation}\label{eq:dyson_expansion}
	\hat{U}(t, t_0) = \hat{U}_1(t, t_0) - i \int_{t_0}^{t} d\tau \hat{U}(t, \tau)\hat{H}_2(\tau)\hat{U}_1(\tau, t_0),
\end{equation}
where $\hat{U}_1$ represents the evolution operator corresponding to $\hat{H}_1$. The generalization of this expression allows flexibility in analyzing perturbative contributions from $\hat{H}_2$.

If we define the system's Hamiltonian components as
\begin{equation}\label{eq:assumed_hamiltonian}
    \hat{H}_1 = \frac{\hat{\mathbf{p}}^2}{2} + V(\mathbf{r}), \quad \hat{H}_2 = V_{le}(\mathbf{r}, t),
\end{equation}
it follows that the transition amplitude can be derived as
\begin{equation}\label{eq:transition_amplitude}
	T(\mathbf{p}) = \lim_{t\to\infty} \langle\psi_{\mathbf{p}}(t)|\psi(t)\rangle.
\end{equation}
Expanding the evolution operator within this framework yields
\begin{eqnarray}
	T(\mathbf{p}) &=& \lim_{t\to\infty, t_0\to-\infty} \Bigg[ \langle\psi_{\mathbf{p}}(t)|\hat{U}_A(t, t_0)|\psi_0(t_0)\rangle \nonumber \\
	&& - i \int_{t_0}^{t} d\tau \langle\psi_{\mathbf{p}}(t)|\hat{U}(t, \tau)V_{le}(\mathbf{r}, \tau)\hat{U}_A(\tau, t_0)|\psi_0(t_0)\rangle \Bigg] \label{eq:expanded_amplitude}.
\end{eqnarray}
Given that the final continuum state $|\psi_{\mathbf{p}}(t)\rangle$ and the initial bound state $|\psi_0(t_0)\rangle$ are eigenstates of $\hat{H}_A$ and orthogonal, the first term vanishes, leading to
\begin{equation}\label{eq:final_result}
	T(\mathbf{p}) = -i \int_{t_0}^{t} d\tau \langle\psi_{\mathbf{p}}(t)|\hat{U}(t, \tau)V_{le}(\mathbf{r}, \tau)|\psi_0(t_0)\rangle.
\end{equation}
Reapplying the Dyson expansion results in:
	\begin{equation}\label{eq:dyson_reapplication}
		T(\mathbf{p}) = -i \int_{t_0}^{t} d\tau \int_{t_0}^{\tau} d\tau' \langle\psi_{\mathbf{p}}(t)|\hat{U}(t, \tau')V(\mathbf{r})\hat{U}_{le}(\tau', \tau)V_{le}(\mathbf{r}, \tau)|\psi_0(t_0)\rangle.
	\end{equation}


\subsection{Strong field approximation}
To simplify the problem within the SFA, the following assumptions are considered:

\begin{enumerate}
	\item The atomic potential $V(\mathbf{r})$ is ignored for continuum state evolution, approximating $\hat{U} \approx \hat{U}_{le}$.
	\item The final state of the electron is assumed to be a free plane wave, given by $\langle r|\psi_{\mathbf{p}}(t)\rangle = (2\pi)^{-3/2} e^{i\mathbf{p} \cdot \mathbf{r} - i \epsilon_p t}$.
	\item The initial bound state $|\psi_0(t_0)\rangle$ does not interact with other bound states under the external field.
\end{enumerate}

Within this approximation, the evolution operator $\hat{U}_{le}$ can be represented in terms of a momentum basis:

\begin{equation}\label{eq:momentum_basis}
	\hat{U}_{le}(t, t') = \int d^3k\ |\chi_{\mathbf{k}}(t)\rangle\langle\chi_{\mathbf{k}}(t')|.
\end{equation}

Together with the assumptions (ii) and (iii), we can express Eqn.\ref{eq:expanded_amplitude} in terms of the following integral form for the transition amplitude \( T(\mathbf{p}) \):

\begin{eqnarray}
	T(\mathbf{p}) &=& \lim_{t \to \infty} \lim_{t_0 \to -\infty} 
	\Bigg[
	-i \int_{t_0}^t \mathrm{d}\tau \int \mathrm{d}^3\mathbf{k} \, \langle \mathbf{p}(t) | \chi_{\mathbf{k}}(t) \rangle 
	\langle \chi_{\mathbf{k}}(\tau) | V_{\mathrm{le}}(\mathbf{r}, \tau) | \psi_0(\tau) \rangle \nonumber \\
	& & + (-i)^2 \int_{t_0}^t \mathrm{d}\tau \int_{\tau}^t \mathrm{d}\tau' \int \mathrm{d}^3\mathbf{k} \, \langle \mathbf{p}(t) | \chi_{\mathbf{k}}(t) \rangle \nonumber \\
	& & \times \langle \chi_{\mathbf{k}}(\tau') | V(\mathbf{r}) U_{\mathrm{le}}(\tau', \tau) V_{\mathrm{le}}(\mathbf{r}, \tau) | \psi_0(\tau) \rangle
	\Bigg].
	\label{eq:transition-amplitude}
\end{eqnarray}

To simplify the expression, we take the limit \( t \to \infty \), using the identity

\begin{equation}
	\lim_{t \to \infty} \langle \mathbf{p}(t) | \chi_{\mathbf{k}}(t) \rangle = e^{i \varphi_\infty} \delta(\mathbf{p} - \mathbf{k}),
	\label{eq:delta-identity}
\end{equation}

where \( \varphi_\infty \) is an arbitrary constant phase factor that is irrelevant to the physical results and can be ignored. Using Eqn.~\ref{eq:delta-identity} in  Eqn.~\ref{eq:transition-amplitude}, the transition amplitude simplifies to the following form

\begin{equation}
	T(\mathbf{p}) = T_0(\mathbf{p}) + T_1(\mathbf{p}),
	\label{eq:sfa-transition-amplitude}
\end{equation}
where the two terms \( T_0(\mathbf{p}) \) and \( T_1(\mathbf{p}) \) are defined as follows:

\begin{equation}
	T_0(\mathbf{p}) = -i \int_{-\infty}^\infty \mathrm{d}\tau \, \langle \chi_{\mathbf{p}}(\tau) | V_{\mathrm{le}}(\mathbf{r}, \tau) | \psi_0(\tau) \rangle,
	\label{eq:t0-term}
\end{equation}

and

\begin{equation}
	T_1(\mathbf{p}) = (-i)^2 \int_{-\infty}^\infty \mathrm{d}\tau \int_{\tau}^\infty \mathrm{d}\tau' \, \langle \chi_{\mathbf{p}}(\tau') | V(\mathbf{r}) U_{\mathrm{le}}(\tau', \tau) V_{\mathrm{le}}(\mathbf{r}, \tau) | \psi_0(\tau) \rangle.
	\label{eq:t1-term}
\end{equation}

Here, \( T_0(\mathbf{p}) \) corresponds to the direct ionization pathway, where the electron is ionized directly from the ground state \( | \psi_0 \rangle \) due to the interaction potential \( V_{\mathrm{le}}(\mathbf{r}, \tau) \). The term \( T_1(\mathbf{p}) \), on the other hand, represents a more complex process, including rescattering effects where the electron interacts with the potential \( V(\mathbf{r}) \) after ionization.

The above equations form the foundation for calculating ATI spectra and photoelectron momentum distributions within the framework of the SFA. This method is particularly useful in describing ionization dynamics induced by intense laser fields.


\subsection{Laser pulse description}
The interaction of a laser field with an atomic or molecular system is described by the laser-electron potential \( V_{le}(\mathbf{r}, t) \). This potential characterizes the interaction of the charged particle with the oscillating electromagnetic field of the laser pulse. The description of the laser-atom interaction can be formulated in different gauges, with the two most commonly used being the length gauge and the velocity gauge.

In the length gauge, the interaction potential is written in terms of the electric field:

\begin{equation}
	V_{le}^{\mathrm{(length)}}(\mathbf{r}, t) = -\mathbf{E}(\mathbf{r}, t) \cdot \mathbf{r},
\end{equation}

where \( \mathbf{E}(\mathbf{r}, t) \) is the time-dependent electric field of the laser pulse, and \( \mathbf{r} \) represents the position vector of the electron. The dipole approximation assumes that the spatial variation of the field over atomic or molecular dimensions is negligible, which holds for long-wavelength laser fields.

In the velocity gauge, the interaction potential is expressed in terms of the vector potential:
\begin{equation}
	V_{le}^{\mathrm{(velocity)}}(\mathbf{p}, t) =\mathbf{A}(\mathbf{r}, t)\cdot\mathbf{p} + \frac{1}{2} \mathbf{A}^2(\mathbf{r}, t),
\end{equation}
where \( \mathbf{p} \) represents the canonical momentum of the electron. The velocity gauge is often preferred in strong-field physics, particularly within the framework of the strong-field approximation (SFA), as it naturally incorporates gauge invariance and facilitates the analysis of momentum distributions.

In theoretical descriptions of laser-atom interactions, the electric field \( \mathbf{E}(\mathbf{r}, t) \) or the vector potential \( \mathbf{A}(\mathbf{r}, t) \) is used. These quantities are related by
\begin{equation}
	\mathbf{E}(\mathbf{r}, t) = -\frac{\partial \mathbf{A}(\mathbf{r}, t)}{\partial t} - \nabla \phi(\mathbf{r}, t).
\end{equation}
In this study, we adopt the velocity gauge, which is advantageous for momentum-space calculations and allows for a more direct interpretation of the photoelectron momentum distributions.
In the Coulomb gauge, where \( \nabla \cdot \mathbf{A} = 0 \), the vector potential \( \mathbf{A}(\mathbf{r}, t) \) satisfies the wave equation
\begin{equation}
	\nabla^2 \mathbf{A} - \frac{1}{c^2} \frac{\partial^2 \mathbf{A}}{\partial t^2} = 0.
\end{equation}
By assuming a monochromatic plane wave solution, we obtain the Helmholtz equation
\begin{equation}
	\nabla^2 \mathbf{A} + k^2 \mathbf{A} = 0,
\end{equation}
where \( k = \frac{\omega_0}{c} \) is the wave number associated with the laser frequency \(\omega_0\).

A general solution to the Helmholtz equation for a plane wave propagating in the direction of the wave vector \(\mathbf{k}\) is given by
\begin{equation}
	\mathbf{A}(\mathbf{r}, t) = \mathbf{A}_0 f(t) e^{i (\mathbf{k} \cdot \mathbf{r} - \omega t + \phi_{\mathrm{cep}})} + \mathrm{c.c.}
\end{equation}
where \( \mathbf{A}_0 \) is the amplitude of the laser pulse, \( f(t) \) is the envelope function describing the temporal profile, \( \phi_{\mathrm{cep}} \) is the carrier-envelope phase (CEP), and \( \omega \) is the central frequency.

To describe elliptical polarization, the vector potential can be decomposed into orthogonal components along the \( x \)- and \( y \)-axes
\begin{equation}
	\mathbf{A}(\mathbf{r}, t) = \frac{A_0}{\sqrt{1 + \epsilon^2}} f(\mathbf{r},t) \left[ \cos(\mathbf{k} \cdot \mathbf{r} - \omega t + \phi_{\mathrm{cep}})\mathbf{e}_x - \epsilon \Lambda \sin(\mathbf{k} \cdot \mathbf{r} - \omega t + \phi_{\mathrm{cep}})\mathbf{e}_y \right],
	\label{eq:vectorPotential}
\end{equation}
where \( A_0 \) is the peak amplitude of the laser pulse, \( \epsilon \) is the ellipticity parameter that determines the polarization state, and \( \Lambda \) represents the helicity with values \( +1 \) for right-handed and \( -1 \) for left-handed polarization. To ensure energy conservation, the normalization factor \( \frac{1}{\sqrt{1 + \epsilon^2}} \) is introduced, which preserves the total intensity. 

The laser pulse is characterized by a sine-squared envelope function, which smoothly modulates the field amplitude to avoid abrupt discontinuities. The envelope function used in this study is defined as
\begin{equation}
	f(\mathbf{r}, t) =
	\left\{
	\begin{array}{ll}
		\sin^2\left(\frac{\mathbf{k} \cdot \mathbf{r} -\omega t}{2 n_p}\right), & 0 \leq t \leq \tau_p, \\[8pt]
		0, & \mathrm{otherwise}.
	\end{array}
	\right.
	\label{Eq:Envelope}
\end{equation}
where \( n_p \) is the number of optical cycles in the pulse and \( \tau_p \) is the total pulse duration. The sine-squared envelope provides a gradual turn-on and turn-off of the field, minimizing spectral artifacts. This form of the vector potential is particularly useful in modeling laser-matter interactions, including ATI and high-harmonic generation (HHG) processes.


\subsection{Dipole approximation}
When an atom is subjected to an intense laser field, certain approximations can be applied to simplify the problem. If the wavelength of the laser, \( \lambda_0 \), is considerably larger than the characteristic atomic dimension, such as the Bohr radius \( a_0 \), the spatial variation of the laser field within the interaction region can be neglected. This assumption leads to the dipole approximation, where only the temporal dependence of the laser field is considered.

Under this approximation, the vector potential (\ref{eq:vectorPotential}) of the laser pulse can be re-written as

\begin{equation}
	\mathbf{A}_0 = \frac{A_0}{\sqrt{1 + \epsilon^2}} f(t) \left[ \cos(\omega t + \phi_{\mathrm{cep}})\mathbf{e}_x + \epsilon \Lambda \sin(\omega t + \phi_{\mathrm{cep}})\mathbf{e}_y \right].
	\label{eq:vectorPotential_Dipole}
\end{equation}

In the dipole approximation, the spatial dependence of the field is ignored, focusing solely on its temporal oscillations. This assumption is valid when the atomic system's size is much smaller than the laser wavelength, ensuring the interaction dynamics are primarily influenced by the time-varying electric field.

In the velocity gauge and within the dipole approximation, the evolution of the quantum state \( \chi(t) \) of an electron subjected to an external electromagnetic field is governed by the time-dependent Schrödinger equation (TDSE), which is expressed as:

\begin{equation}
	i\hbar \frac{\partial}{\partial t} \chi(t) = \left[ -\frac{1}{2} \nabla^2 - \mathbf{A}(t) \cdot \nabla + \frac{1}{2} \mathbf{A}^2(t) \right] \chi(t).
\end{equation}

To facilitate further analysis, the wave function is represented in momentum space by introducing the Fourier transform as 

\begin{equation}
	\chi(\mathbf{p}, t) = \langle \mathbf{p} | \chi(t) \rangle.
\end{equation}

Substituting this transformation into the TDSE leads to the modified equation

\begin{equation}
	i\hbar \frac{\partial}{\partial t} \chi(\mathbf{p}, t) = \left[ \frac{\mathbf{p}^2}{2} + \mathbf{A}(t) \cdot \mathbf{p} + \frac{1}{2} \mathbf{A}^2(t) \right] \chi(\mathbf{p}, t).
\end{equation}

Applying the method of separation of variables and integrating over time, the solution in momentum representation takes the exponential form as

\begin{equation}
	\chi(\mathbf{p}, t) = \exp \left( -\frac{i}{2} \int_{0}^{t} \left[ \mathbf{p} + \mathbf{A}(\tau) \right]^2 d\tau \right).
\end{equation}

By performing the inverse Fourier transform, the wave function in position representation is obtained

\begin{equation}
	\chi_{\mathbf{p}}(\mathbf{r}, t) = C e^{i\mathbf{p} \cdot \mathbf{r} - \frac{i}{2} \int_{0}^{t} |\mathbf{p} + \mathbf{A}(\tau)|^2 d\tau},
\end{equation}

where \( C \) is a normalization constant. To determine its value, the normalization condition is applied

\begin{equation}
	\langle \chi_{\mathbf{p}}(\mathbf{r}, t) | \chi_{\mathbf{p'}}(\mathbf{r}, t) \rangle = \delta(\mathbf{p} - \mathbf{p'}).
\end{equation}

This condition leads to the conclusion that \( C = (2\pi)^{-3/2} \), resulting in the final expression for the Volkov wave function under the velocity gauge and dipole approximation:

\begin{equation}
	\chi_{\mathbf{p}}(\mathbf{r}, t) = (2\pi)^{-3/2} e^{i\mathbf{p} \cdot \mathbf{r}} e^{-\dot\iota S(\mathbf{p}, t)},
	\label{Eq:VolkovState}
\end{equation}
where the phase factor also know as the Volkov phase is given by
\begin{equation}
	S(\mathbf{p}, t) =  \frac{1}{2} \int_{0}^{t} (\mathbf{p} + \mathbf{A}(\tau))^2 d\tau.
	\label{Eq:VolkovPhase}
\end{equation}
The Volkov phase describes the classical action of the free electron moving in the laser field.

To solve the Volkov phase, we begin by expressing the vector potential of a elliptically polarized pulse in a more compact form. Starting from the vector potential given in Eq.~\ref{eq:vectorPotential_Dipole}, we rewrite it as \cite{Dar2024,Milošević_2006}

\begin{equation}
	\mathbf{A}(t) = \sum_{j=0}^{2} \frac{\mathcal{A}_j}{\sqrt{1+\epsilon^2}} 
	\left[ \cos(\omega_j t + \phi_{\mathrm{CEP}}) \mathbf{e}_x 
	+ \epsilon\Lambda \sin(\omega_j t + \phi_{\mathrm{CEP}}) \mathbf{e}_y \right],
	\label{eq:VectorPotentialCompact}
\end{equation}

where the trigonometric products have been expanded. Here, the index \( j \) distinguishes between the lower (\( j=0 \)), central (\( j=1 \)), and upper (\( j=2 \)) frequency components of the pulse. The corresponding frequencies are defined as \( \omega_{0} = \left( 1 - \frac{1}{n_p} \right) \omega \), \( \omega_1 = \omega \), and \( \omega_2 = \left( 1 + \frac{1}{n_p} \right) \omega \), where \( \omega \) is the central frequency of the pulse and \( n_p \) represents the number of optical cycles in the pulse. The amplitudes of the vector potential components vary with frequency and are given by \( \mathcal{A}_{0} = -A_0/4 \), \( \mathcal{A}_1 = A_0/2 \), and \( \mathcal{A}_2 = -A_0/4 \), where \( A_0 \) is the peak amplitude of the vector potential. These expressions define the spectral characteristics of the field and its amplitude distribution across the frequency components. The lower (\( j=0 \)) and upper (\( j=2 \)) components have equal magnitudes but opposite signs, while the central component (\( j=1 \)) dominates with twice the amplitude of the side components.

\begin{figure*}
	\centering
	\includegraphics[width=1.0\textwidth]{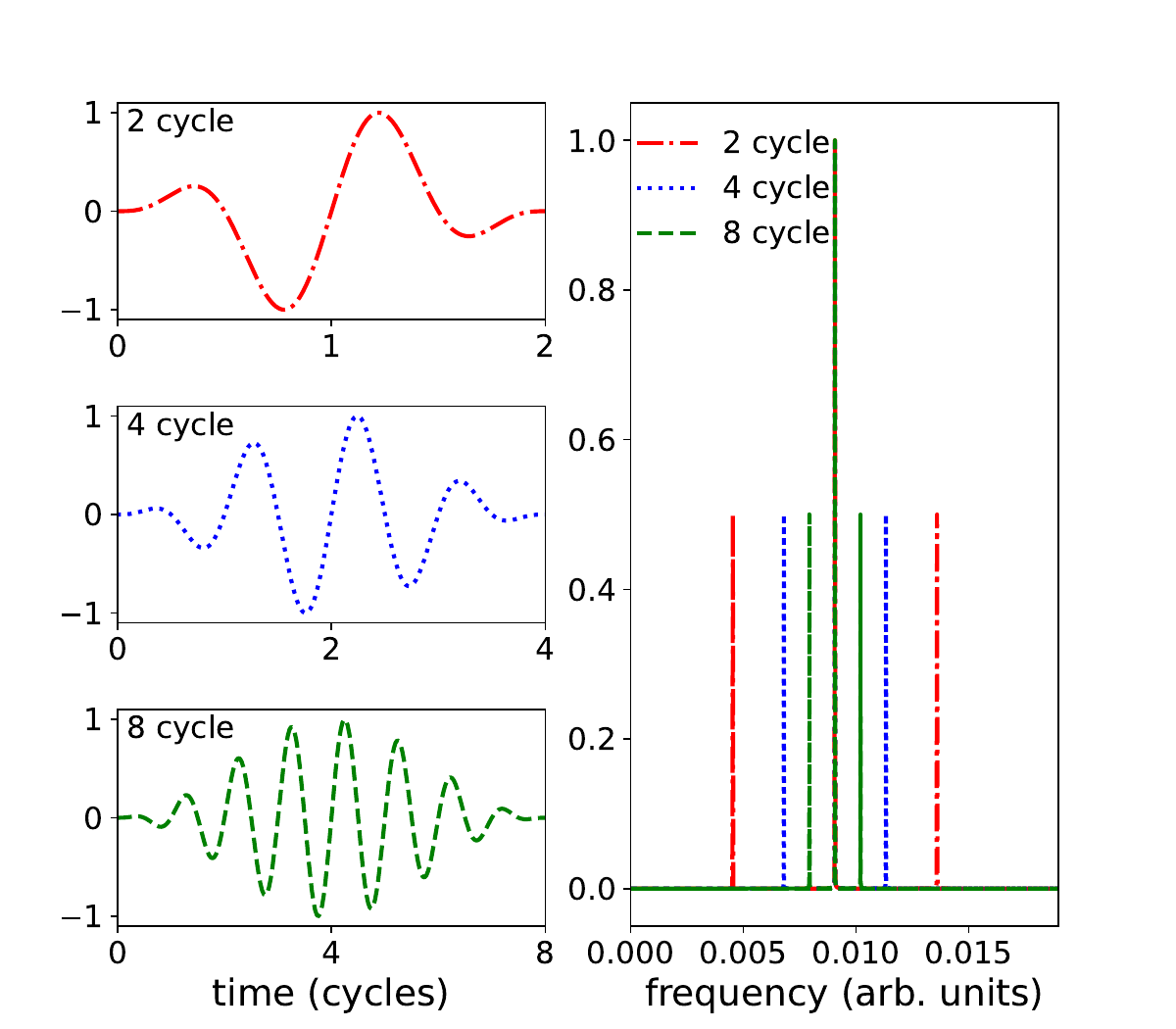}
	\caption{\label{Fig_Pulse}Time-domain representation of the vector potential \( \boldsymbol{A}(t) \) for a circularly polarized laser pulse (\ref{eq:VectorPotentialCompact}) is shown in the left panels, while in the right panel its corresponding frequency-domain representation, \( [\mathcal{F}(\boldsymbol{A})](\omega) \). The plots include laser pulses with durations of two (red), four (blue), and eight (green) optical cycles, with a carrier-envelope phase of \( \varphi_{\mathrm{cep}} = 0 \) and a wavelength of 800 nm. The right panel’s vertical axis represents the absolute amplitude of the vector potential at a peak intensity of \( 5 \times 10^{14} \) W/cm\(^2\).  
		
	}
\end{figure*}

The corresponding vector potential, as well as its Fourier spectra for different optical cycles, is illustrated in Fig.~\ref{Fig_Pulse}. The figure provides a visual representation of the temporal and spectral properties of the pulse, highlighting the contributions of the lower, central, and upper frequency components to the overall structure of the vector potential.

Using this vector potential we carry the integration in the volkov phase and we simply get
\begin{eqnarray}
	S(\mathbf{p}, t) &=& \frac{1}{2} \mathbf{p}^2 t
	+ \frac{t}{4} \sum_{i=0}^{2} \mathcal{A}_i^2 
	+ \frac{1 - \epsilon^2}{1 + \epsilon^2} \sum_{i=0}^{2} \frac{\mathcal{A}_i^2}{8\omega_i} \sin(2\omega_i t + 2\phi_{\mathrm{CEP}}) \nonumber \\
	&& + \sum_{i=0}^{1} \sum_{j=i+1}^{2} \frac{\mathcal{A}_i \mathcal{A}_j}{2 (\omega_i - \omega_j)} \sin((\omega_i - \omega_j)t) \nonumber \\
	&& + \frac{1 - \epsilon^2}{1 + \epsilon^2} \sum_{i=0}^{1} \sum_{j=i+1}^{2} \frac{\mathcal{A}_i \mathcal{A}_j}{2 (\omega_i + \omega_j)} \sin((\omega_i + \omega_j)t + 2\phi_{\mathrm{CEP}}) \nonumber \\
	&& + \frac{p_x}{\sqrt{1 + \epsilon^2}} \sum_{i=0}^{2} \frac{\mathcal{A}_i}{\omega_i} \sin(\omega_i t + \phi_{\mathrm{CEP}}) \nonumber \\
	&& - \epsilon \Lambda \frac{p_y}{\sqrt{1 + \epsilon^2}} \sum_{i=0}^{2} \frac{\mathcal{A}_i}{\omega_i} \cos(\omega_i t + \phi_{\mathrm{CEP}}).
	\label{Eq:VolkovPhaseSol}
\end{eqnarray}

To determine the probability of ionization process, we can refer to Eq.\ref{eq:t0-term}. Given that the operator \( \hat{V}_{le} = \hat{H}_{le} - \hat{H}_A + V(\mathbf{r}) \), the direct amplitude can be further simplified by integrating by parts. Since the vector potential remains nonzero over the interval \( t_i \leq t \leq t_f \), we utilize the operator \( \hat{V}_{le} \) to transform Eq.\ref{eq:t0-term} into a more convenient expression as
\begin{eqnarray}
	T_0(\mathbf{p}) &=& -i \int_{t_i}^{t_f} d\tau \left( \langle \chi_p (\tau) | \overleftarrow{\frac{\partial}{\partial \tau}} + \overrightarrow{\frac{\partial}{\partial \tau}} | \Psi_0 (\tau) \rangle \right) \nonumber \\
	& & - i \int_{t_i}^{t_f} d\tau \langle \chi_p (\tau) | V(\mathbf{r}) | \Psi_0 (\tau) \rangle \nonumber \\
	&=& - \langle \chi_p (\tau) | \Psi_0 (\tau) \rangle \Big|_{t_i}^{t_f}  - i \int_{t_i}^{t_f} d\tau \langle \chi_p (\tau) | V(\mathbf{r}) | \Psi_0 (\tau) \rangle.
	\label{Eq:DirectAmplitudeParts}
\end{eqnarray}
This reformulation simplifies the evaluation of the direct amplitude, making it easier to interpret in terms of time evolution and interaction potential. For a hydrogen-like \( 1s \) wave function, the initial bound state can be effectively approximated using a modified ionization potential \( I_p \). This allows us to express the time-dependent wave function as
\begin{equation}
	|\Psi_0 (t) \rangle = |\Psi_0 \rangle e^{i I_p t} = \frac{(2I_p)^{3/2}}{\sqrt{\pi}} e^{-\sqrt{2 I_p} r} e^{i I_p t}.
	\label{Eq:initialWavefunction}
\end{equation}
This expression represents the bound state, where the wave function maintains an exponential decay in space and an oscillatory phase factor in time due to the binding energy \( I_p \).

Together with Eq. \ref{Eq:VolkovState} and Eq. \ref{Eq:initialWavefunction} we can rewrite the direct transition amplitude as
\begin{equation}
	T_0(\mathbf{p}) = - \langle \mathbf{p} | \Psi_0 \rangle e^{-\dot\iota (S(\mathbf{p}, \tau)+ I_p \tau)} \Big|_{t_i}^{t_f}  - \dot\iota \langle \mathbf{p} | V(\mathbf{r}) | \Psi_0 \rangle\int_{t_i}^{t_f} d\tau e^{-\dot\iota (S(\mathbf{p}, \tau)+ I_p \tau)}.
	\label{Eq:DirectTransitionAmplitude}
\end{equation}
The matrix element of the Coulomb potential and the initial wave function in momentum space can be expressed as 
\begin{equation}
	\langle \mathbf{p} | V(\mathbf{r}) | \Psi_0 \rangle = - \frac{2^{\frac{3}{2}} I_p^{\frac{5}{4}}}{\pi} \frac{1}{\epsilon_p + I_p}.
\end{equation}
Similarly, the initial wave function in momentum space is given by:
\begin{equation}
	\langle \mathbf{p} | \Psi_0 \rangle = \frac{(2 I_p)^{\frac{5}{4}}}{\pi \sqrt{2}} \frac{1}{\left( \epsilon_p + I_p \right)^2}.
\end{equation}

To evaluate the integral in Eq.~\ref{Eq:DirectTransitionAmplitude}, we must account for the oscillatory nature of the Volkov phase \( S(\mathbf{p}, \tau) \), given in Eq.~\ref{Eq:VolkovPhaseSol}. The presence of trigonometric terms in \( S(\mathbf{p}, \tau) \) leads to rapidly oscillating exponentials in the transition amplitude, making direct integration challenging. To simplify this integral, we employ the Jacobi-Anger expansion, which allows us to express exponentials terms as infinite series of Bessel functions:

\begin{equation}
	e^{\pm\dot\iota z \sin\theta} = \sum_{n=-\infty}^{\infty} J_n(z) e^{\pm\dot\iota n \theta} \qquad  \mathrm{and} \qquad e^{\pm\dot\iota z \cos\theta} = \sum_{n=-\infty}^{\infty} (\pm\dot \iota)^n J_n(z) e^{\dot\iota n \theta}
\end{equation}

Applying this expansion to the oscillatory terms in the Volkov phase, we can rewrite the exponential term in Eq.~\ref{Eq:DirectTransitionAmplitude} as a summation over discrete harmonics, each associated with a Bessel function \( J_n(z) \). This transformation decomposes the electron-laser interaction into separate photon absorption channels, where the index \( n \) represents the number of absorbed photons.

This expansion dissolves the trigonometric functions in the exponent and makes it easier to carry the integration. Thus after applying the integration within the duration of the pulse \([0,\tau_p]\) we arrive at the final expression of the direct transition amplitude as
\begin{eqnarray}
	T^D_0(\mathbf{p}) = \prod_{i=1}^{15} \left\{ (-\dot\iota)^{n_i \Theta(i-12)} \sum_{n_i=-\infty}^{\infty} J_{n_i} (x_i^D) \right. 
	&&  \left. \left( \frac{\langle \mathbf{p} | V(\mathbf{r}) | \Psi_0 \rangle}{\epsilon_p + U_p + \Omega_{n_i}^D + I_p} + \langle \mathbf{p} | \Psi_0 \rangle \right) \right. \nonumber\\ 
	&& \left. \times \left( 1 - e^{\dot\iota \left[\left(\epsilon_p + U_p + \Omega_{n_i}^D +I_p\right) \tau_p + \Phi_i^D\right]} \right) \right\},
	\label{Eq:TransitionAmplitudeJacobi}
\end{eqnarray}
where \( \Theta(x) \) is the Heaviside step function which ensures that the extra factor \( i^{n_i} \) appears only for the last three terms (\( i \geq 13 \)), while it remains absent for the first 12 terms (\( i \leq 12 \)). Here we redefined the frequencies and phases in terms of \(\Omega\) and \(\Phi\) respectively and including \(x_i\) they are defined in the Appendix \ref{Appendix}. The superscript D is introduced to distinguish this transition amplitude from ones that will be derived later in the nondipole case. The pondermotive energy \(U_p\) is given by \(U_p = \frac{\mathcal{A}_j}{4}\) for \(j = 0:2\).  To approximate the expansion coefficients effectively, the argument of each Bessel function imposes a natural cut-off for the corresponding index \( n_i \). Beyond this limit, the Bessel function exhibits an exponential decline, following the relation \( J_{n_i}(x) \sim e^{-n_i} \) for values where \( |n_i| > |x| = n_{i,\max} \). Consequently, the fifteen originally infinite summations can be truncated within the range \( -n_{i,\max} \leq n_i \leq n_{i,\max} \), providing a reliable approximation.


\subsection{Saddle point method}

In the framework of the SFA, one of the fundamental tasks is to evaluate the time-dependent transition amplitude Eqn. \ref{eq:t0-term}.
In previous subsection, the Jacobi-Anger expansion was employed to solve the transition amplitude by expressing the Volkov phase as a series of Bessel functions. Although this approach provides an analytical representation of the solution, it has significant limitations in practical applications. The main drawbacks include the slow convergence of the infinite series, which leads to computational inefficiencies, and the difficulty in accurately capturing the rapid oscillatory behavior of the wave functions involved. As a result, an alternative and more efficient approach is required to overcome these challenges.

Once computed, this integral provides access to key physical observables related to the ionization dynamics of an atomic or molecular system subjected to an intense laser field. However, the direct numerical evaluation of this integral poses significant computational challenges due to the presence of a rapidly oscillating exponential term. Specifically, the transition amplitude contains an exponential phase term governed by the quasi-classical action. The most significant contributions arise from points in complex time where the phase is stationary. This requirement leads to the saddle-point equation, which is obtained by differentiating the action function.

Since the action is given by the integral
\begin{equation}
	S(\mathbf{p}, t) = \int^{t} d\tau \left[ \frac{1}{2} (\mathbf{p} + \mathbf{A}(\tau))^2 + I_p \right],
\end{equation}
the saddle points are determined by differentiating with respect to $t$, leading to the stationary phase condition
\begin{equation}
	\frac{dS}{dt} = \left[ \frac{1}{2} (\mathbf{p} + \mathbf{A}(t_s))^2 + I_p \right] = 0.
\end{equation}

This equation represents an energy conservation condition at complex ionization times, stating that the kinetic energy of the outgoing electron, combined with the ionization potential, must be zero at the moment of ionization. Solving this equation yields a set of complex ionization times $t_s$, which correspond to the instants of ionization.

Since these times are complex-valued, they are expressed as
\begin{equation}
	t_s = t_r + i t_i,
\end{equation}
where $t_r$ represents the real part (corresponding to classical ionization time), and $t_i$ is the imaginary part, which is directly related to the tunneling probability. The imaginary component signifies the tunneling delay, governing how long the electron remains under the influence of the potential barrier before escaping.

To evaluate the integral in Eqn.~\ref{eq:t0-term} efficiently, the real-time is deformed into the complex plane. The new integration path is chosen such that the oscillations in the exponent are minimized. This is achieved by ensuring that along the steepest descent path, the real part of the exponent decays rapidly while the imaginary part remains stationary.
\begin{figure*}
	\centering
	\includegraphics[width=0.8\textwidth]{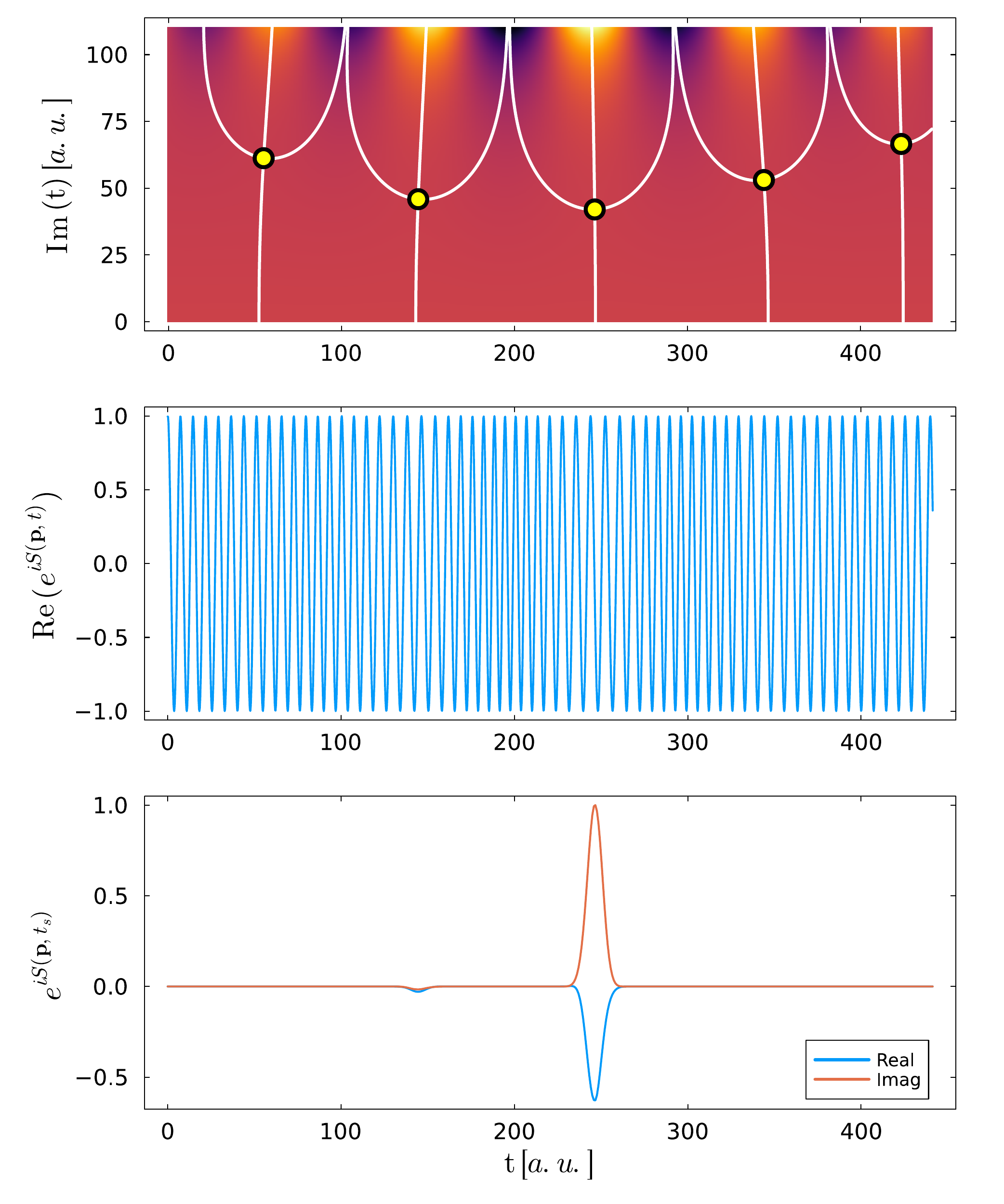}
	\caption{\label{Fig_SaddlePoint}
			Visualization of the saddle points, Volkov phase, and contour integration for a four-cycle circularly polarized laser pulse.  
			(Top) The complex-time saddle points (yellow dots) and the corresponding contour of integration (white curves) are overlaid on a density plot of the action’s imaginary part.  
			(Middle) The real part of the Volkov phase, \( e^{iS(\boldsymbol{p}, t)} \), is plotted as a function of real time, showing rapid oscillations.  
			(Bottom) The Volkov phase evaluated along the integration contour or the steepest-descent path, \( e^{iS(\boldsymbol{p}, t_s)} \), where \( t_s \) represents the complex saddle-point time. The integration smooths the oscillatory behavior, highlighting the dominant contributions to the ionization amplitude.  
	}
\end{figure*}
To provide a formal and detailed description of the role of saddle points in strong-field ionization, we begin by expanding the action function \( S(\mathbf{p}, t) \) around each saddle point \( t_s \) using a Taylor series expansion. This expansion is truncated at the second order, yielding the following approximation:

\begin{equation}
	S(\mathbf{p}, t) \approx S(\mathbf{p}, t_s) + \frac{1}{2} S''(\mathbf{p}, t_s) (t - t_s)^2,
\end{equation}

where \( S(\mathbf{p}, t_s) \) represents the value of the action at the saddle point \( t_s \), and \( S''(\mathbf{p}, t_s) \) denotes the second derivative of the action with respect to time, evaluated at \( t_s \). The second derivative of the action is explicitly given by:

\begin{equation}
	S''(\mathbf{p}, t_s) = -\mathbf{E}(t_s) \cdot \left[ \mathbf{p} + \mathbf{A}(t_s) \right],
\end{equation}

where \( \mathbf{E}(t) = -\dot{\mathbf{A}}(t) \) is the laser electric field, and \( \mathbf{A}(t) \) is the vector potential of the laser field. This expression highlights the dependence of the curvature of the action on both the laser electric field and the canonical momentum of the electron, \( \mathbf{p} + \mathbf{A}(t_s) \).

Given the highly oscillatory nature of the exponential term \( e^{i S(\mathbf{p},t)} \), the time integral over this term is approximated using the method of Gaussian integration. Specifically, the integral of interest is:

\begin{equation}
	\int dt \, e^{i S''(t_s) (t - t_s)^2 / 2}.
\end{equation}

By applying the standard Gaussian integral formula, this integral evaluates to:

\begin{equation}
	\int dt \, e^{i S''(t_s) (t - t_s)^2 / 2} = \sqrt{\frac{2 \pi i}{S''(t_s)}}.
\end{equation}

The resulting prefactor \( \sqrt{\frac{2 \pi i}{S''(t_s)}} \) plays a critical role in determining the amplitude of the ionization probability, as it encapsulates the influence of the saddle point on the ionization dynamics.

Fig. \ref{Fig_SaddlePoint} present a graphical representation of the saddle points in the complex time plane for a four-cycle circularly polarized laser pulse (800 nm, $5 \times 10^{14}$ W/cm$^2$), with an electron momentum of \( 0.5 \) atomic units (a.u.) for an Argon atom. The figure is divided into three panels. The top panel displays the imaginary part of the ionization time trajectory, with white contour lines indicating the steepest descent paths used in the evaluation of the integral. These paths are essential for ensuring the convergence of the Gaussian approximation. The middle panel presents the real part of \( e^{iS(\mathbf{p},t)} \), revealing rapid oscillations due to the influence of the laser electric field. These oscillations underscore the complexity of the ionization process and the necessity of a robust theoretical framework to accurately describe it. The bottom panel emphasizes the dominant contribution of the saddle points to the ionization process. The concentration of the integral's weight around these points reinforces the importance of the Gaussian approximation in providing an accurate theoretical treatment of strong-field ionization.

With the integral now approximated in terms of dominant saddle-point contributions, the final expression for the transition amplitude follows
\begin{equation}
	T^{SP}_0(\mathbf{p}) = \langle \mathbf{p} | V(\mathbf{r}) | \Psi_0 \rangle \sum_{t_s} \sqrt{\frac{2 \pi i}{\mathbf{E}(t_s) \cdot (\mathbf{p} + \mathbf{A}(t_s))}} e^{\dot \iota (S(\mathbf{p}, t_s) + I_p t_s)}.
	\label{Eq:TransitionAmplitudeSaddle}
\end{equation}
This result encapsulates the dominant quantum pathways of ionization, with each term in the sum corresponding to a different saddle-point solution. The prefactor accounts for the laser field strength at the moment of ionization, while the exponential term captures the essential quantum-mechanical phase associated with the electron's transition from the bound state to the continuum.


\subsection{Non-dipole approximation}
To incorporate the nondipole effects within the SFA, it is essential to express the vector potential of the driving laser field as a function of position, $\mathbf{r}$. This implies that the vector potential exhibits both spatial and temporal dependencies, leading to the emergence of both electric and magnetic field components. Such a formulation is crucial for an accurate representation of the laser field and its influence on electron motion within the strong-field regime. When a laser field propagates with an angular frequency $\omega$ along the wave vector direction $\mathbf{k}=\frac{\omega}{c}\hat{k}$, the vector potential can be formulated as a sum of plane-wave components
\begin{eqnarray}
		\mathbf{A}(\mathbf{r},t) = \int d^{3}\mathbf{k} \mathbf{A} (\mathbf{k},t) = \mathrm{Re}\{\mathbf{a} (\mathbf{k}) e^{\dot{\iota} (\mathbf{k} \cdot \mathbf{r} - \omega_{\mathbf{k}} t)}\}.
		\label{eq:Fourier_vector_potential}
\end{eqnarray}
Here, $\mathbf{a} (\mathbf{k})$ represents the complex Fourier coefficient of the vector potential. This formalism allows us to describe the continuum state solution as presented in Ref.~\cite{Boning2019}, given by

\begin{equation}
	\chi_{\mathbf{p}} (\mathbf{r},t) = \frac{1}{(2\pi)^{\frac{3}{2}}} e^{-\dot{\iota} (\epsilon_{p} t - \mathbf{p} \cdot \mathbf{r})} e^{-\dot{\iota} \Gamma(\mathbf{r},t)},
	\label{eq:nondipole_volkov_state}
\end{equation}
which corresponds to a modified Volkov state. This state explicitly incorporates the particle's momentum $\mathbf{p}$, along with the influence of the electromagnetic field through the modified Volkov phase $\Gamma(\mathbf{r},t)$. The interaction dynamics between the electron and the field are governed by the classical Lorentz force equation, with the overall wavefunction expressed as a coherent sum over plane-wave components, each characterized by distinct momentum and energy values.

The influence of the electromagnetic field on the particle's motion is embedded within the Volkov state through the phase factors of the plane waves. This allows for an analytical description of the electron’s trajectory and energy evolution in the presence of an external field. The modified Volkov phase, as presented in Ref.~\cite{boning2019}, can be expressed as
\begin{eqnarray}
			\Gamma(\mathbf{r},t) &=& \int d^{3}\mathbf{k} \rho_{\mathbf{k}} \sin(\mathrm{u}_{\mathbf{k}} + \theta_{\mathbf{k}}) \nonumber \\
			&& + \int d^{3}\mathbf{k} \int d^{3}\mathbf{k}' \Big[ \alpha^{+}_{\mathbf{k},\mathbf{k}'} \sin(\mathrm{u}_{\mathbf{k}} + \mathrm{u}_{\mathbf{k}'} + \theta^{+}_{\mathbf{k},\mathbf{k}'}) 
			+ \alpha^{-}_{\mathbf{k},\mathbf{k}'} \sin(\mathrm{u}_{\mathbf{k}} - \mathrm{u}_{\mathbf{k}'} + \theta^{-}_{\mathbf{k},\mathbf{k}'}) \Big]\nonumber \\
			&& + \frac{1}{2} \int d^{3}\mathbf{k} \int d^{3}\mathbf{k}' \sigma_{\mathbf{k},\mathbf{k}'} \rho_{\mathbf{k}} 
			\left( \frac{\sin(\mathrm{u}_{\mathbf{k}} + \mathrm{u}_{\mathbf{k}'} + \theta_{\mathbf{k}} + \xi_{\mathbf{k},\mathbf{k}'})}{\eta(\mathbf{k}) + \eta(\mathbf{k}')} 
			+ \frac{\sin(\mathrm{u}_{\mathbf{k}} - \mathrm{u}_{\mathbf{k}'} + \theta_{\mathbf{k}} - \xi_{\mathbf{k},\mathbf{k}'})}{\eta(\mathbf{k}) - \eta(\mathbf{k}')} \right).\nonumber\\
			\label{eq:modified_volkov_phase}
\end{eqnarray}
In this formulation, $\rho_{\mathbf{k}}, \theta_{\mathbf{k}}, \sigma_{\mathbf{k},\mathbf{k}'}$, and $\xi_{\mathbf{k},\mathbf{k}'}$ serve as projection operators, which are explicitly tied to the Fourier components of the vector potential, $\mathbf{a} (\mathbf{k})$, and the electron's momentum, $\mathbf{p}$. These operators define how $\mathbf{p}$ and the wave vector $\mathbf{k}$ project onto the momentum-space representation of the field $\mathbf{A}(\mathbf{k}',t)$. Furthermore, the terms $\alpha^{\pm}_{\mathbf{k},\mathbf{k}'}$ characterize the ponderomotive contributions corresponding to each mode, being proportional to the inner product of the Fourier coefficients, $[\mathbf{a} (\mathbf{k}) \cdot \mathbf{a} (\mathbf{k}')]$. Detailed definitions of these operators can be found in Appendix \ref{Appendix}, where we also introduce $\eta(\mathbf{k}) = \mathbf{p} \cdot \mathbf{k} - \omega_{\mathbf{k}}$.

Similar to the vector potential in Eq.~\ref{eq:VectorPotentialCompact}, the vector potential for a few-cycle pulse with a spatial dependence can also be written as a sum of plane waves 
\begin{equation}
	\mathbf{A}(\mathbf{r}, t) = \sum_{j=0}^{2} \frac{\mathcal{A}_j}{\sqrt{1 + \epsilon^2}} 
	\left[ \cos(u_j + \phi_{\mathrm{cep}}) \mathbf{e}_x
	+ \epsilon \Lambda \sin(u_j + \phi_{\mathrm{cep}}) \mathbf{e}_y \right],
	\label{Eq:nondipoleVactorPotential}
\end{equation}
where we use the short notation \( u_j = \mathbf{k}_j \cdot \mathbf{r} - \omega_j t \). Using this vector potential the modified Volkov phase can be solved by carrying the integration as
\begin{eqnarray}
	\Gamma(\mathbf{r}, t) &=& \sum_{j=0}^{2}\frac{\mathcal{A}_j^2}{4}\frac{u_j}{\eta_j(\mathbf{k})} + \frac{1-\epsilon^2}{1+\epsilon^2} \sum_{j=0}^{2} \frac{\mathcal{A}_j^2}{8\eta_j(\mathbf{k})} \sin(2u_j + 2\phi_{\mathrm{cep}}) \nonumber\\
	&+& \sum_{i=0}^{1} \sum_{j=i+1}^{2} \frac{\mathcal{A}_i\mathcal{A}_j}{2(\eta_i(\mathbf{k}) -\eta_j(\mathbf{k}))} \sin(u_i - u_j)\nonumber \\
	&+& \frac{1-\epsilon^2}{1+\epsilon^2} \sum_{i=0}^{1} \sum_{j=i+1}^{2} \frac{\mathcal{A}_i\mathcal{A}_j}{2(\eta_i(\mathbf{k}) +\eta_j(\mathbf{k}))} \sin(u_i + u_j + 2\phi_{\mathrm{cep}})\nonumber\\
	&+& \frac{p_x}{\sqrt{1 + \epsilon^2}} \sum_{j=0}^{2} \frac{\mathcal{A}_j}{\eta_j(\mathbf{k})} \sin(u_j + \phi_{\mathrm{CEP}}) \nonumber \\
	&-& \epsilon \Lambda \frac{p_y}{\sqrt{1 + \epsilon^2}} \sum_{j=0}^{2} \frac{\mathcal{A}_j}{\eta_j(\mathbf{k})} \cos(u_j + \phi_{\mathrm{CEP}}).
	\label{Eq:VolkovPhaseSolNondipole}
\end{eqnarray}
The obtained solution closely resembles the one derived in the dipole approximation (\ref{Eq:VolkovPhaseSol}), with the primary distinction being the additional spatial dependence introduced through the terms \( u_j \) and \( \eta_j(\mathbf{k}) \). These terms account for nondipole effects, modifying the transition amplitude accordingly.
To further simplify the expression and eliminate the sine and cosine terms present in the exponent, we employ the Jacobi-Anger expansion. This allows us to rewrite the oscillatory terms in a more tractable form, facilitating the subsequent integration.
By following this approach, we obtain the complete solution for the transition amplitude in the nondipole case, expressed as
\begin{eqnarray}
	T^{ND}_0(\mathbf{p}) = \prod_{i=1}^{15} \left\{ \sum_{n_i-\infty}^{\infty} (-\dot\iota)^{n_i \Theta(i-12)} J_{n_i} (x_i^{ND}) \right. 
	&& \left. \left( \frac{\langle \mathbf{p}_{n_i} | V(\mathbf{r}) | \Psi_0 \rangle}{\epsilon_p + U_p + \Omega_{n_i}^{ND} +I_p} + \langle \mathbf{p}_{n_i} | \Psi_0 \rangle \right) \right. \nonumber \\
	&& \left. \times \left( 1 - e^{\dot\iota \left[\left(\epsilon_p + U_p + \Omega_{n_i}^{ND} +I_p\right) \tau_p + \Phi_{n_i}^{ND}\right]} \right) \right\}.
	\label{Eq:TransitionAmplitudeNondipole}
\end{eqnarray}
as before we have \( \Theta(x) \) as the Heaviside step function which ensures that the extra factor \( i^{n_i} \) appears only for the last three terms (\( i \geq 13 \)), while it remains absent for the first 12 terms (\( i \leq 12 \)). Similarly, we redefined the frequencies and phases in terms of \(\Omega\) and \(\Phi\) respectively and including \(x_i\) and \(\mathbf{p}_i\) are defined in the Appendix \ref{Appendix}. Similarly, the pondermotive energy \(U_p\) is given by \(U_p = \sum_j\frac{\mathcal{A}_j}{4\eta_j}\) for \(j = 0:2\).

\subsection{Nondipole Saddle point method}
Following the derivation of the nondipole Volkov state in the previous subsection, which incorporates the magnetic field component of the laser interaction, we now focus on applying the saddle point method to solve the resulting complex integral expressions. This approach requires careful separation of the purely time-dependent contribution of the Volkov phase from the contributions that depend on both time and spatial coordinates. To achieve this, we employ a Taylor expansion of the relevant trigonometric functions around \(\mathbf{k} \cdot \mathbf{r} = 0\).

To perform the Taylor expansion of a function of the form \(\sin(\mathbf{k} \cdot \mathbf{r} - \omega t + \phi)\) around \(\mathbf{k} \cdot \mathbf{r} = 0\), we first expand the argument of the sine function and then the sine function itself. Let us define the argument as:
\[
f \approx -(\omega t - \phi) + (\mathbf{k} \cdot \mathbf{r}),
\]
where the expansion is performed around \(\mathbf{k} \cdot \mathbf{r} = 0\). Next, we expand \(\sin(f)\) around \(f = 0\) using the Taylor series expansion for \(\sin(x)\) about \(x = 0\):
\[
\sin(x) \approx x - \frac{x^3}{6} + \frac{x^5}{120} - \ldots.
\]
Substituting \(f\) into this expansion, we obtain:
\[
\sin(f) \approx \sin(-\omega t + \phi) + \cos(-\omega t + \phi) (\mathbf{k} \cdot \mathbf{r}) - \frac{\sin(-\omega t + \phi)}{6} (\mathbf{k} \cdot \mathbf{r})^2 + \ldots
\]
This expansion can be carried out either before or after the momentum integration without compromising accuracy. Consequently, the nondipole Volkov phase can be rewritten in a separable form as \cite{Minneker2022}:
\begin{equation}
	\Gamma(\mathbf{r}, t) = \Gamma_1(t) + \mathbf{r} \cdot \mathbf{\Gamma}_2(t),
	\label{eq:volkov_phase_separation}
\end{equation}
where \(\mathbf{\Gamma}_2(t) = -\partial_t \Gamma_1(t) \frac{\mathbf{k}}{\omega_{\mathbf{k}}}\). Using this separation, the nondipole Volkov state can be expressed as:
\begin{equation}
	\chi_{\mathbf{p}} (\mathbf{r},t) = \frac{1}{(2\pi)^{\frac{3}{2}}} e^{i \left[ (\mathbf{p} - \mathbf{\Gamma}_2(t)) \cdot \mathbf{r}\right]} e^{-i (\epsilon_{p} t + \Gamma_1(t))}.
	\label{eq:nondipole_volkov_state2}
\end{equation}

Using the vector potential defined in Eq.~\ref{Eq:nondipoleVactorPotential}, the temporal part of the phase, \(\Gamma_1(t)\), can be evaluated as 
\begin{eqnarray}
	\Gamma_1(t) &=& \sum_{j=0}^{2} \frac{\mathcal{A}_j^2}{4} \frac{-\omega_j t}{\eta_j(\mathbf{k})} + \frac{1-\epsilon^2}{1+\epsilon^2} \sum_{j=0}^{2} \frac{\mathcal{A}_j^2}{8\eta_j(\mathbf{k})} \sin(-2\omega_j t + 2\phi_{\mathrm{cep}}) \nonumber\\
	&+& \sum_{i=0}^{1} \sum_{j=i+1}^{2} \frac{\mathcal{A}_i\mathcal{A}_j}{2(\eta_i(\mathbf{k}) -\eta_j(\mathbf{k}))} \sin(-(\omega_i - \omega_j)t) \nonumber \\
	&+& \frac{1-\epsilon^2}{1+\epsilon^2} \sum_{i=0}^{1} \sum_{j=i+1}^{2} \frac{\mathcal{A}_i\mathcal{A}_j}{2(\eta_i(\mathbf{k}) +\eta_j(\mathbf{k}))} \sin(-(\omega_i + \omega_j) t + 2\phi_{\mathrm{cep}}) \nonumber\\
	&+& \sqrt{\frac{p_x^2 + \epsilon^2 p_y^2}{1 + \epsilon^2}} \sum_{j=0}^{2} \frac{\mathcal{A}_j}{\eta_j(\mathbf{k})} \sin(-\omega_j t + \phi_{\mathrm{cep}} - \varphi_p^{(\epsilon)}),
	\label{eq:gamma1_temporal_phase}
\end{eqnarray}
where \(\varphi_p = \arctan\left(\frac{\epsilon p_y}{p_x}\right)\).

With the integral now approximated in terms of dominant saddle-point contributions, the final expression for the nondipole transition amplitude can be written as
\begin{equation}
	T^{NDSP}_0(\mathbf{p}) = \sum_{t_s} \langle \mathbf{p} - \mathbf{\Gamma}_2(t_s) | V(\mathbf{r}) | \Psi_0 \rangle \sqrt{\frac{2 \pi i}{\partial^2_{t_s}\Gamma_1(t_s)}} e^{i (\epsilon_{p} t_s + \Gamma_1(t_s) + I_p t_s)},
	\label{Eq:TransitionAmplitudeSaddle2}
\end{equation}
where the superscript NDSP denotes the nondipole saddle point transition amplitude. This expression highlights the role of the saddle points \(t_s\) in determining the dominant contributions to the transition amplitude.
\section{Results and Discussion}

\subsection{Photoelectron Momentum Distributions (PMD)}
\begin{figure*}
	\centering
	\includegraphics[width=1.0\textwidth]{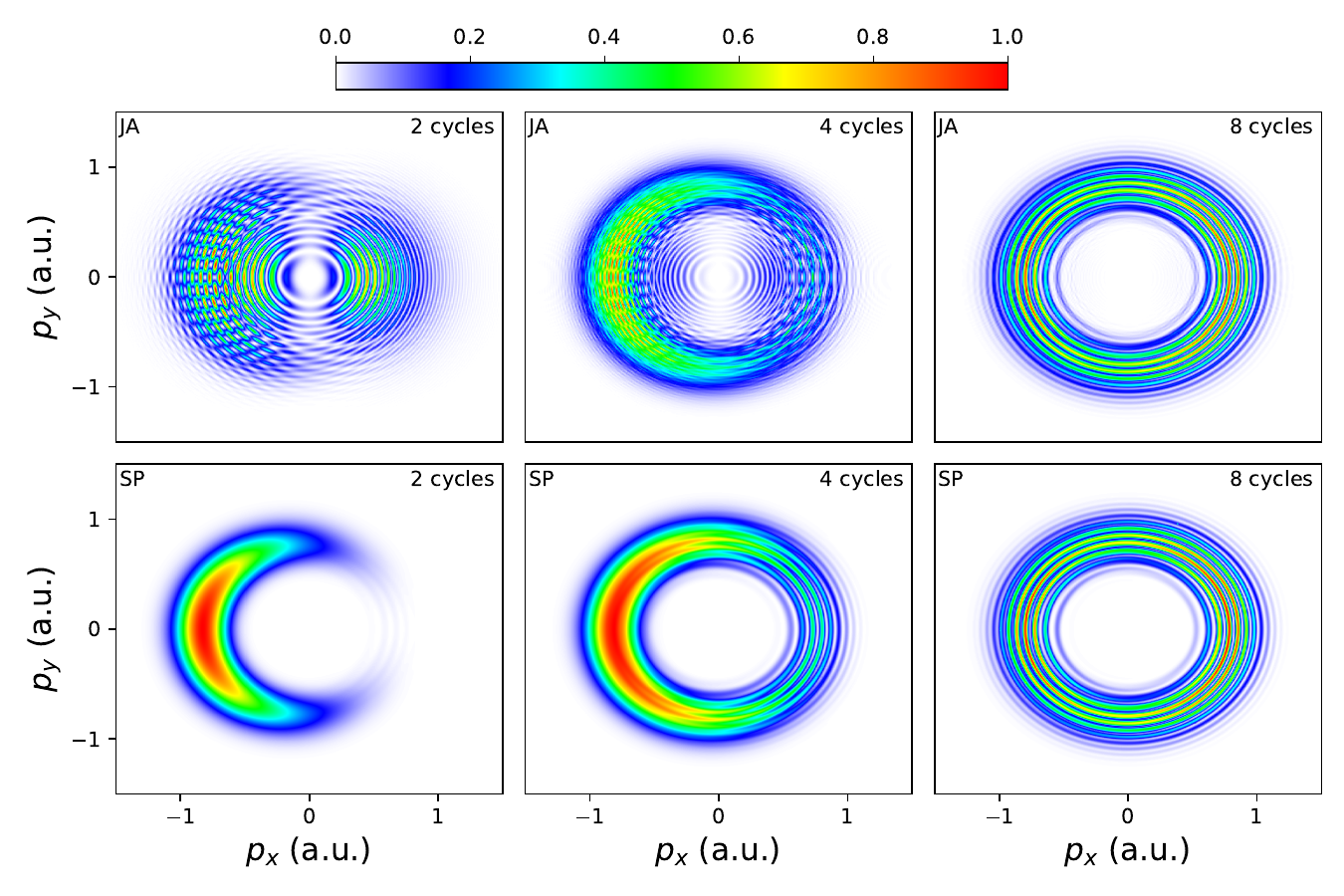}
	\caption{\label{Fig_PMD}Photoelectron momentum distributions in the laser polarization plane for ionization of an argon atom by a circularly polarized laser pulse with a wavelength of 800 nm and peak intensity of \( 5 \times 10^{14} \) W/cm\(^2\).  
		The top row (JA) presents results obtained using the Jacobi-Anger expansion, while the bottom row (SP) corresponds to the saddle-point method. Each column represents different pulse durations: two-cycle (left), four-cycle (middle), and eight-cycle (right) pulses. The color scale represents the normalized probability amplitude.  
	}
\end{figure*}

The photoelectron momentum distributions (PMDs) resulting from the strong-field ionization of argon atoms by a circularly polarized laser pulse are analyzed using two distinct theoretical approaches: the Jacobi-Anger (JA) expansion and the saddle-point (SP) method. The results, presented in Fig.~\ref{Fig_PMD}, highlight the capabilities of these methods in capturing the key features of ionization dynamics for varying pulse durations.

The observed interference patterns in the PMDs are closely linked to the spectral composition of the laser pulse \cite{Martiny2007}. As illustrated in Fig.~\ref{Fig_Pulse}, the time-domain representation of the pulse for different cycle durations reveals the presence of multiple frequency components. The corresponding power spectra demonstrate that shorter pulses exhibit a broader range of contributing frequencies, which enhances interference effects in the ionization process.

For a two-cycle pulse, the frequency spectrum is notably broad, leading to a large number of ionization pathways with nearly equal contributions. These pathways interfere coherently, producing intricate oscillatory structures in the momentum distributions. As the pulse duration increases to four and eight cycles, the frequency spectrum narrows, and the interference effects become more regular, forming well-defined concentric rings.

The JA expansion (top row of Fig.~\ref{Fig_PMD}) reveals fine oscillatory structures in the PMDs, particularly for the two-cycle pulse. These high-frequency oscillations arise from the coherent superposition of multiple ionization pathways, reflecting the interference effects induced by the broad spectral content of short pulses. As the pulse duration increases to four and eight cycles, the structures evolve into more concentric patterns, with well-defined circular interference fringes. This trend indicates that longer pulses lead to more stable ionization dynamics, reducing the transient spectral effects observed in shorter pulses.

In contrast, the SP method (bottom row of Fig.~\ref{Fig_PMD}) provides a complementary perspective by focusing on the dominant ionization pathways. Unlike the JA expansion, which accounts for all interfering ionization paths, the SP method selects only the most significant saddle points. Consequently, it captures the basic structure of the photoelectron distribution without resolving finer interference patterns. For the two-cycle pulse, the distribution appears broader and lacks the detailed oscillations seen in the JA approach. As the pulse duration increases, the SP results exhibit clearer ring-like structures, reflecting the primary energy contributions from ionization events at specific phases of the laser cycle.

A key observation is the increasing prominence of circular structures in both methods as the pulse duration grows. For the two-cycle pulse, the JA method reveals strong interference features, while the SP approach yields a smooth, broad distribution without capturing high-order oscillations. This difference arises because the SP method focuses only on the dominant saddle points, neglecting weaker pathways that contribute to interference.

At four cycles, the interference fringes in the JA method become more organized, forming clearer concentric patterns. The SP method continues to provide a simplified structure that approximates the primary energy distribution without resolving detailed interference effects. In the eight-cycle case, both methods produce a well-defined ring structure, though the JA method retains weak interference features at low photoelectron energies.

The comparison between the JA expansion and the SP method underscores the importance of selecting the appropriate theoretical framework based on the application. The JA expansion effectively captures all interfering ionization pathways, making it particularly suitable for analyzing fine-scale structures in ATI spectra. On the other hand, the SP method offers a computationally efficient approach to estimate the dominant photoelectron distributions by considering only the most significant ionization pathways. This trade-off between accuracy and computational cost highlights the complementary strengths of the two methods.

\subsection{Above-Threshold Ionization (ATI) Spectra}

\begin{figure*}
	\centering
	\includegraphics[width=1.0\textwidth]{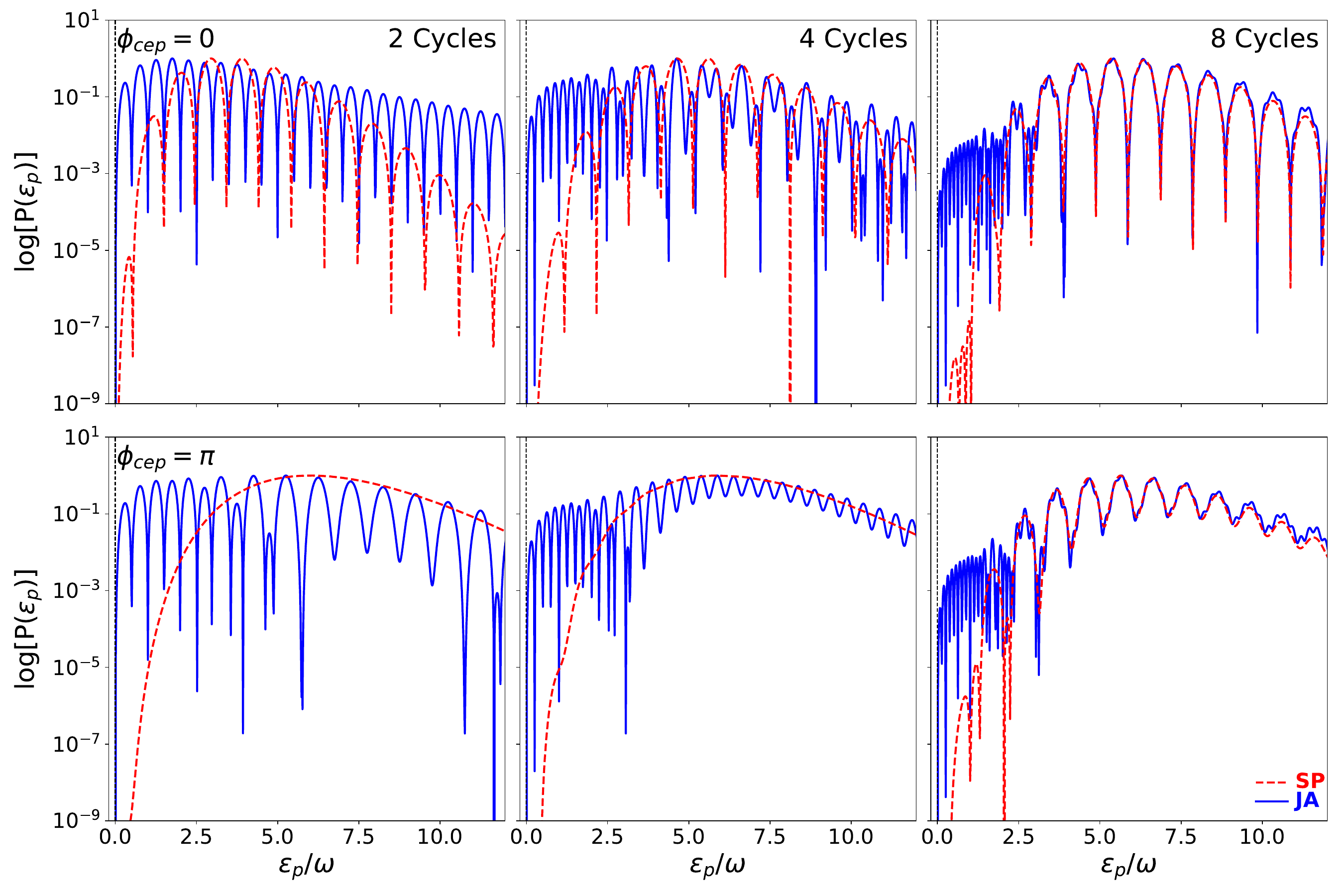}
	\caption{\label{Fig_ATI}Normalized ATI spectra comparing the saddle-point method (SP) (red dashed lines) and Jacobi-Anger expansion (JA) (blue solid lines) for an argon atom ionized by a laser pulse with a wavelength of 800 nm and peak intensity of \( 5 \times 10^{14} \) W/cm\(^2\).  
		The spectra are presented as a function of scaled photoelectron energy \( \varepsilon_p / \omega \), with different pulse durations: 2 cycles (left), 4 cycles (middle), and 8 cycles (right). The upper row corresponds to a carrier-envelope phase of \( \varphi_{\mathrm{cep}} = 0 \), while the lower row shows results for \( \varphi_{\mathrm{cep}} = \pi \). The laser propagation direction is chosen such that the polar angle is \( 90^\circ \) and the azimuthal angle is \( 0^\circ \). 
	}
\end{figure*}

Figure~\ref{Fig_ATI} presents the ATI spectra as a function of scaled photoelectron energy ($\varepsilon_p / \omega$), comparing the results obtained from the SP method and the JA expansion for different pulse durations and CEPs. The top row corresponds to $\varphi_{\mathrm{cep}} = 0$, while the bottom row shows results for $\varphi_{\mathrm{cep}} = \pi$. Each column represents a different pulse duration: two-cycle (left), four-cycle (middle), and eight-cycle (right).

The ATI spectra reveal nonlinear interference effects, which arise from the interaction of ionized electrons with the frequency components embedded in the laser pulse. These interference patterns result from the coherent superposition of multiple ionization pathways contributing to the final photoelectron energy distribution.

Shorter laser pulses, characterized by a broader frequency spectrum, introduce a wider range of ionization pathways. As seen in the JA results (solid blue lines), this leads to stronger interference structures, with prominent oscillations appearing in the spectra. For the two-cycle pulse, the ATI spectrum exhibits a fine structure with 0.5 $\hbar \omega$ peak separation, a feature that is well-captured by the JA method but absent in the SP results. This finer structure originates from interference between ionization pathways arising from the frequency components of the electric field, which the SP method does not account for, as it primarily considers only the dominant ionization trajectories. This behavior aligns with our previous findings \cite{Dar2024}.

As the pulse duration increases to four and eight cycles, the frequency spectrum narrows, and the contributions become increasingly dominated by specific energy channels. Consequently, the ATI spectra in the JA method evolve into more regular and well-defined peaks, with reduced high-frequency oscillations. This trend reflects the transition from transient spectral effects in shorter pulses to more stable ionization dynamics in longer pulses.

\begin{figure*}
	\centering
	\includegraphics[width=1.0\textwidth]{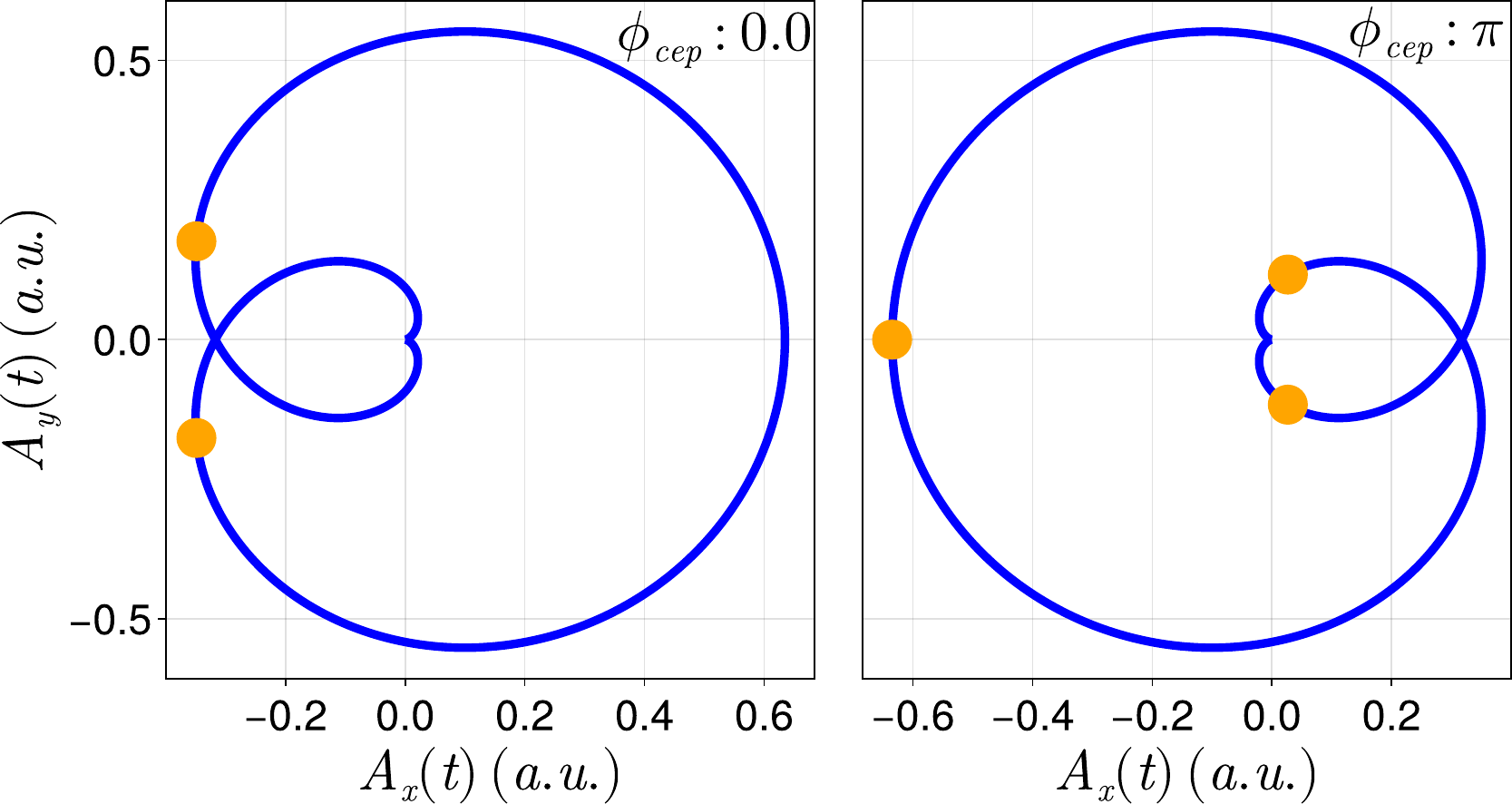}
	\caption{\label{Fig_saddlepointxy}Vector potential for a two-cycle laser pulse at a wavelength of 800 nm, with a peak intensity of $5 \times 10^{14} \, \mathrm{W/cm}^2$ and an ionization potential $I_p = 15.7596 \, \mathrm{eV}$. The left panel shows the vector potential for a $\phi_{\mathrm{CEP}}$ of 0, and the right panel for $\phi_{\mathrm{CEP}} = \pi$. The yellow dots indicate the positions of saddle-point solutions relevant to the ionization events. 
	}
\end{figure*}

The CEP plays a crucial role in shaping the interference effects in the ATI spectra. Comparing the two CEP cases ($\varphi_{\mathrm{cep}} = 0$ and $\varphi_{\mathrm{cep}} = \pi$), it is evident that the phase shifts modify the relative contributions of different ionization pathways, leading to variations in the spectral shape. For $\varphi_{\mathrm{cep}} = 0$, the oscillatory structure in the JA results is more pronounced at lower energies, while for $\varphi_{\mathrm{cep}} = \pi$, a shift in the interference pattern occurs, altering the spectral intensity distribution. These CEP-dependent variations highlight the sensitivity of the ionization process to the temporal structure of the laser pulse. In Fig. \ref{Fig_saddlepointxy}, it is evident that for a carrier-envelope phase (CEP) of zero, the dominant contribution to the temporal integral arises from two saddle points. In contrast, for a CEP of \(\pi\), the integral is primarily influenced by a single saddle point. This observation provides a plausible explanation for the persistence of oscillations at zero CEP in the case of a two-cycle pulse. 

The SP method (dashed red lines) provides a complementary perspective by focusing on the dominant ionization pathways. Unlike the JA expansion, which accounts for all interfering ionization paths, the SP method selects only the most significant saddle points. As a result, it captures the fundamental structure of the ATI spectrum without resolving the fine-scale interference effects. For the two-cycle pulse, the SP results exhibit a smooth envelope that approximates the overall energy distribution but lacks the detailed oscillations seen in the JA approach. As the pulse duration increases, the SP method continues to reproduce the primary energy peaks, albeit without the high-frequency interference structures.  

\subsection{Role of Ellipticity}

\begin{figure*}
	\centering
	\includegraphics[width=0.8\textwidth]{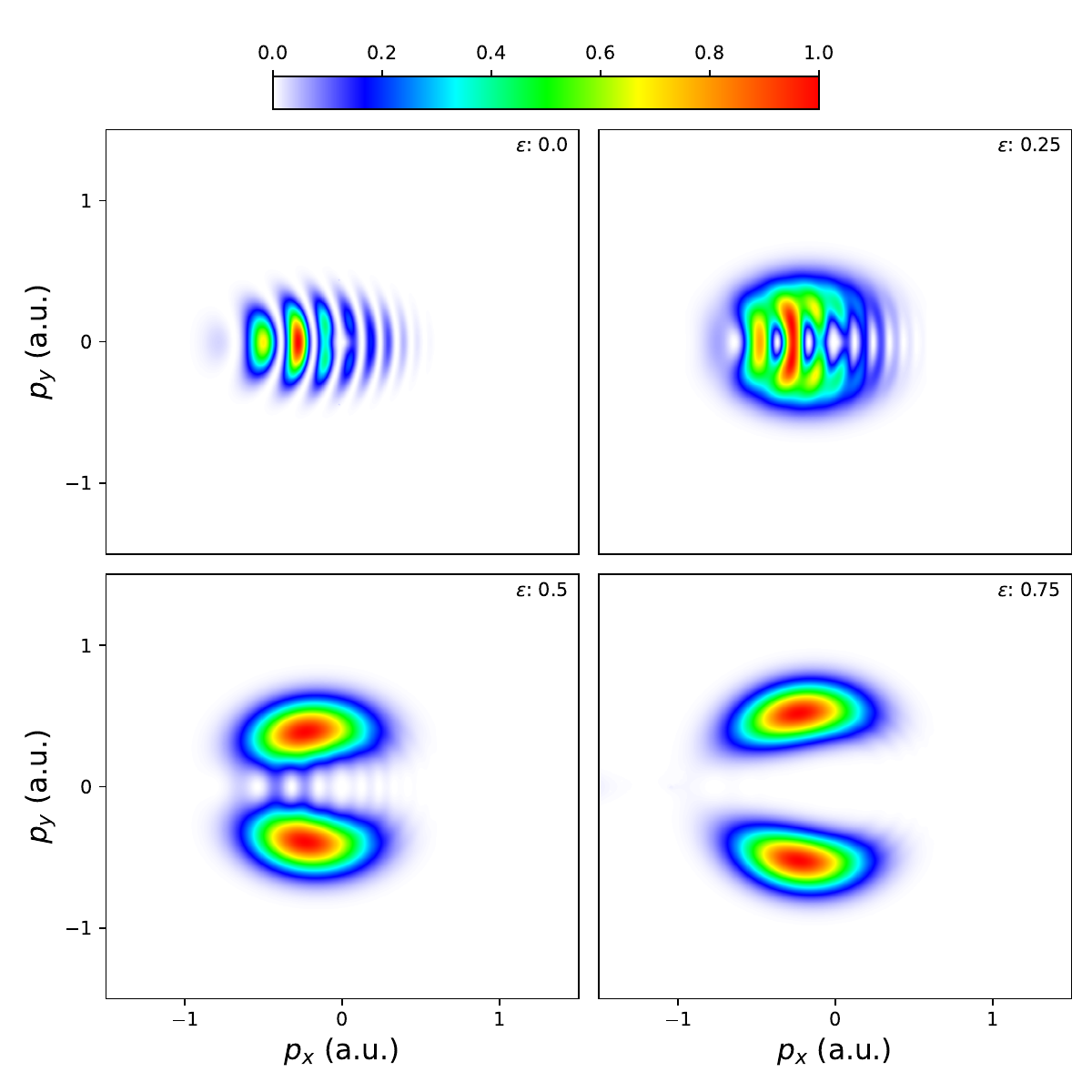}
	\caption{\label{Fig_Ellipticity}Photoelectron momentum distributions in the laser polarization plane for ionization of an argon atom by a two-cycle laser pulse with a wavelength of 800 nm and peak intensity of \( 5 \times 10^{14} \) W/cm\(^2\).  
		The ellipticity \(\varepsilon\) of the laser field is varied from \( \varepsilon = 0.0 \) (top-left) to \( \varepsilon = 0.75 \) (bottom-right). The color scale represents the normalized probability amplitude.  
	}
\end{figure*}

Figure~\ref{Fig_Ellipticity} presents the PMD in the polarization plane $(p_x, p_y)$ for a two-cycle laser pulse with varying ellipticity $\varepsilon$. The ellipticity increases from $\varepsilon = 0.0$ (top-left) to $\varepsilon = 0.75$ (bottom-right), demonstrating the evolution of the electron emission pattern as the polarization transitions from linear to elliptical. The laser parameters are consistent with previous figures, featuring a wavelength of $800$ nm and a peak intensity of $5 \times 10^{14}$ W/cm$^2$, ionizing an argon atom. The short pulse duration (two cycles) ensures that the ionization process is dominated by a small number of optical cycles, allowing for a clear observation of the interplay between the laser field and electron dynamics. The saddle point method is employed to analyze the ionization process, providing a semiclassical interpretation of the electron trajectories and their contributions to the final momentum distribution.

\begin{figure*}
	\centering
	\includegraphics[width=1.0\textwidth]{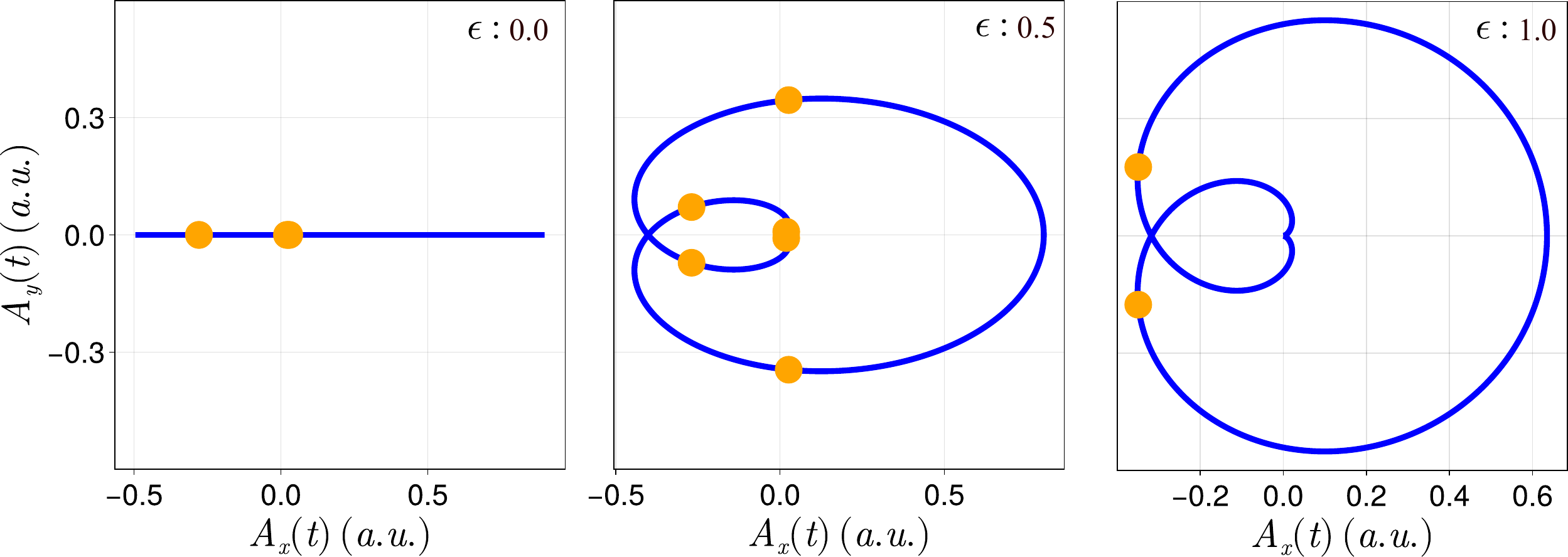}
	\caption{\label{Fig_saddlepointeps}Vector potential for a two-cycle laser pulse at a fixed $\phi_{\mathrm{CEP}} = 0$ while varying the ellipticity $\epsilon$. The panels show ellipticity values of $\epsilon = 0.0$, $\epsilon = 0.5$, and $\epsilon = 1.0$, respectively. The yellow dots represent saddle-point solutions, indicating critical points for ionization that shift with changes in the ellipticity. The other parameters are same as Fig. \ref{Fig_saddlepointxy}
	}
\end{figure*}

For linear polarization ($\varepsilon = 0.0$, top-left panel), the momentum distribution exhibits well-defined interference fringes. These fringes arise from the constructive and destructive interference of electron wave packets ionized at different times within the two-cycle pulse. The saddle point method, as seen in Fig. \ref{Fig_saddlepointeps}, reveals that these ionization events correspond to distinct release times, each associated with a specific phase of the laser field. The distribution is highly asymmetric, with dominant electron emission along the major polarization axis ($p_x$-axis), reflecting the alignment of the laser's electric field. The multiple fringes are a signature of the coherent superposition of electron trajectories, where the phase differences between saddle points lead to interference patterns in the momentum distribution.

As the ellipticity increases to $\varepsilon = 0.25$ (top-right panel), the momentum distribution broadens, and the interference fringes become more complex. The photoelectron emission begins to spread along the minor polarization axis ($p_y$-axis), indicating the growing influence of the elliptical component of the laser field. The saddle point method shows that the additional momentum components along the $y$-axis arise from the non-zero electric field in this direction, which modifies the classical trajectories of the electrons. Despite this, the interference pattern remains visible, suggesting that multiple ionization pathways continue to contribute to the final momentum distribution. The persistence of interference fringes at this ellipticity highlights the robustness of quantum coherence in the ionization process, even in the presence of an elliptical field.

At $\varepsilon = 0.5$ (bottom-left panel), a significant transformation occurs. The distribution splits into two distinct lobes, symmetrically centered around the origin. This bifurcation is a hallmark of elliptical polarization, where the electron emission is no longer confined to a single axis but instead follows the polarization ellipse. The saddle point method reveals that the two lobes correspond to electrons ionized at different phases of the laser field, see Fig. \ref{Fig_saddlepointeps}, with each lobe associated with a specific set of saddle points. The interference fringes along the minor axis become less pronounced, and the primary emission aligns with the polarization ellipse. This transition from a structured interference pattern to a smoother, more localized distribution is a direct consequence of the reduced coherence between ionization pathways as the ellipticity increases. The reduced coherence arises because the elliptical field introduces a time-dependent phase shift between ionization events, disrupting the strict phase relationships observed at lower ellipticities.

For higher ellipticity ($\varepsilon = 0.75$, bottom-right panel), the photoelectron momentum distribution adopts a clear two-lobed structure, with emission predominantly along the major axis of the elliptical polarization. The interference fringes vanish entirely, and the distribution aligns with the classical expectation for ionization in an elliptically polarized field. The saddle point method confirms that the absence of structured interference at this ellipticity is due to the dominance of direct ionization pathways, where the electron trajectories are less likely to overlap. 

\begin{figure*}
	\centering
	\includegraphics[width=1.0\textwidth]{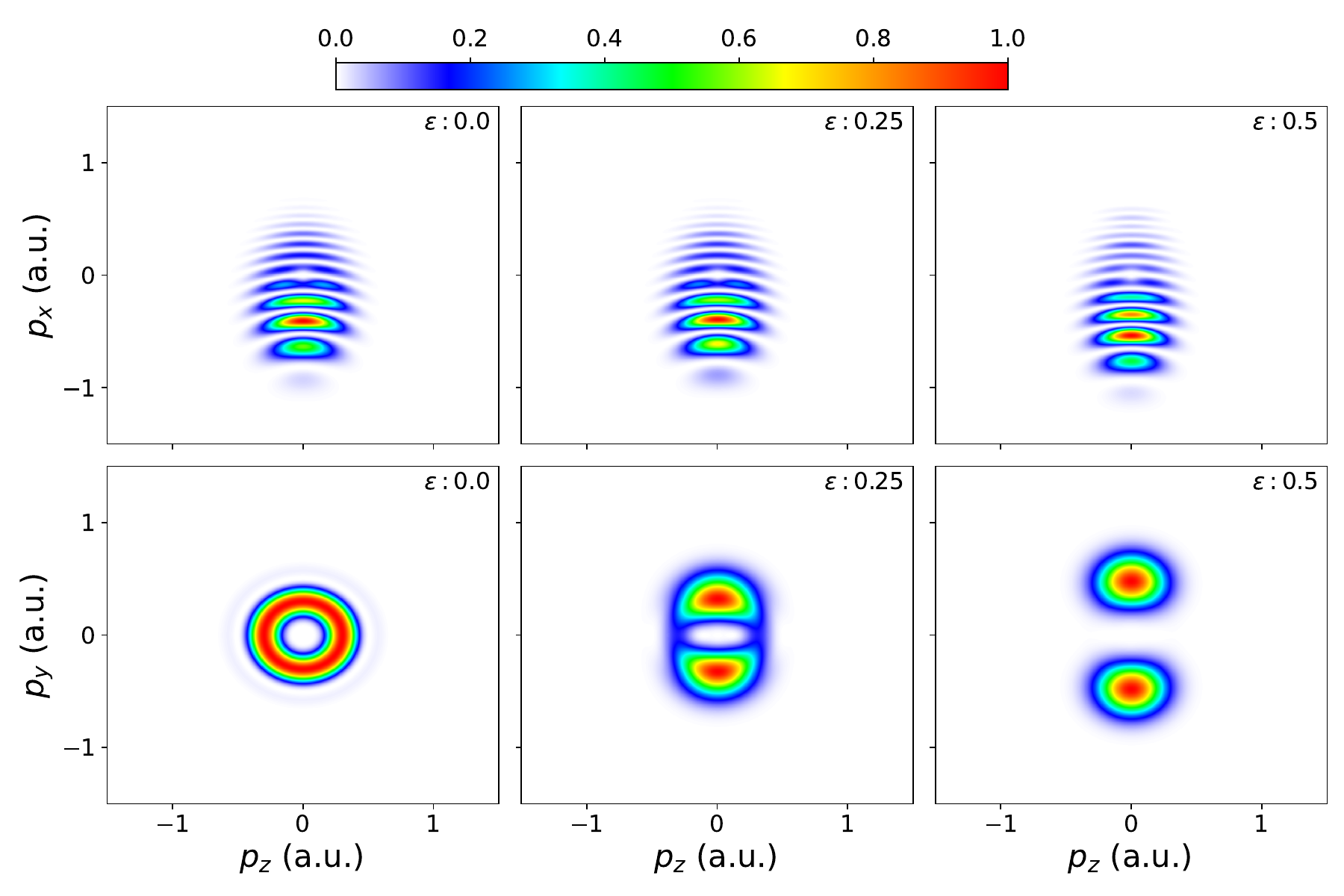}
	\caption{\label{Fig_polarization}Photoelectron momentum distributions in the laser propagation plane for ionization of an argon atom by a two-cycle laser pulse with a wavelength of 800 nm and peak intensity of \( 5 \times 10^{14} \) W/cm\(^2\).  
		The distributions are shown in the \( (p_x, p_z) \) plane (top row) and the \( (p_y, p_z) \) plane (bottom row), where the ellipticity \(\varepsilon\) of the laser field is varied from \( \varepsilon = 0.0 \) (left) to \( \varepsilon = 0.5 \) (right). The color scale represents the normalized probability amplitude.  
	}
\end{figure*}

Figure~\ref{Fig_polarization} presents the PMD in the laser propagation plane, illustrating electron emission patterns in the $(p_x, p_z)$ and $(p_y, p_z)$ planes for a two-cycle laser pulse with varying ellipticity $\varepsilon$. The ellipticity increases from $\varepsilon = 0.0$ (left) to $\varepsilon = 0.5$ (right), providing insights into how electron dynamics are influenced by elliptical polarization.

For linear polarization ($\varepsilon = 0.0$), the momentum distributions exhibit well-defined interference fringes, particularly in the $(p_x, p_z)$ plane. These fringes result from the coherent superposition of multiple ionization pathways, as predicted by the saddle point method. The distribution in the $(p_y, p_z)$ plane retains a concentric circular shape, reflecting the strong symmetry in the emission pattern due to the absence of elliptical field components. The circular symmetry in the $(p_y, p_z)$ plane is a direct consequence of the linear polarization, where the electric field oscillates purely along the $x$-axis, leaving the $y$-axis unaffected.

As the ellipticity increases to $\varepsilon = 0.25$, the interference fringes in the $(p_x, p_z)$ plane remain visible but begin to lose contrast. The electron emission becomes more localized around the central axis, and the distribution in the $(p_y, p_z)$ plane starts to elongate along the polarization direction, reflecting the influence of the elliptical component of the laser field. The saddle point method shows that this elongation is due to the additional momentum components introduced by the non-zero electric field along the $y$-axis. The transition from concentric rings to a more stretched distribution highlights the increasing role of ellipticity in modifying electron trajectories.

At $\varepsilon = 0.5$, the distributions undergo significant changes. In the $(p_x, p_z)$ plane, the fringes are strongly suppressed, resulting in a smooth and elongated structure. This suppression arises from the reduced coherence between different ionization events, as the elliptical field introduces additional momentum components that disrupt the strict phase relationships observed at lower ellipticities. In the $(p_y, p_z)$ plane, the momentum distribution splits into two distinct lobes, aligned along the major axis of the elliptical polarization.

\subsection{Non-Dipole Effects in Strong-Field Ionization}
\begin{figure*}
	\centering
	\includegraphics[width=1.0\textwidth]{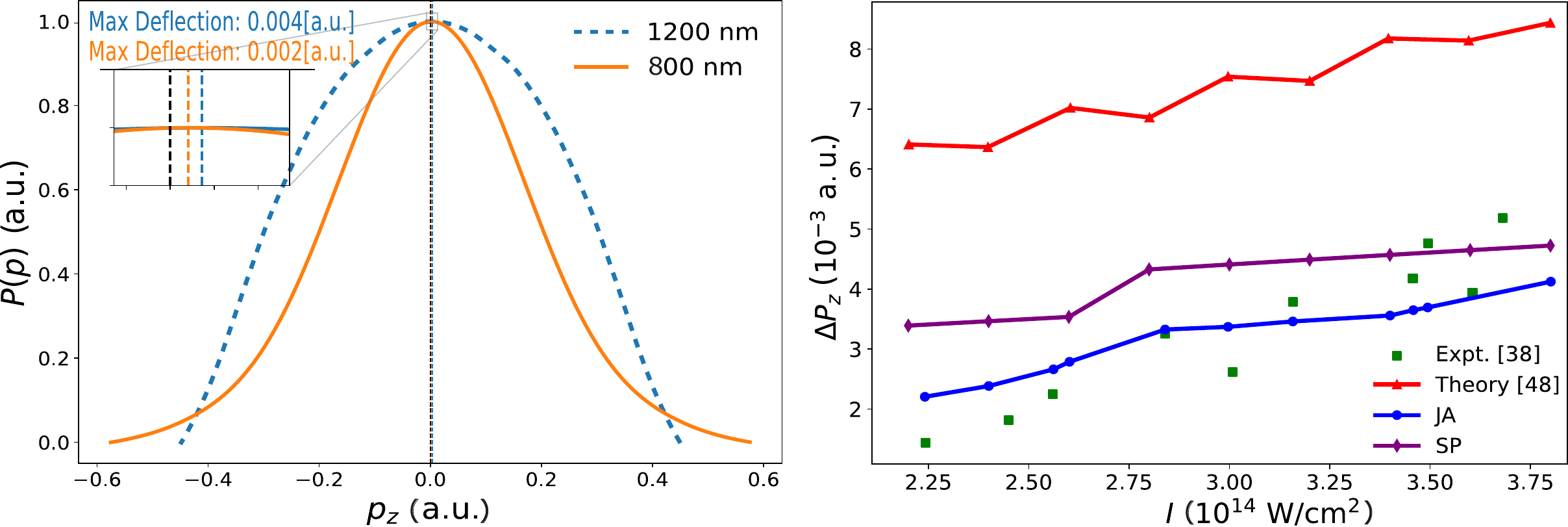}
	\caption{\label{Fig_NDshift}\textbf{Left:} Ionization probability \( P(\mathbf{p}) \) driven by circularly polarized, 2-cycle laser pulses at 800 nm (solid orange) and 1200 nm (dashed blue), with a peak intensity of \( 5 \times 10^{14} \, \mathrm{W/cm}^2 \). The distributions exhibit a wavelength-dependent shift along the propagation direction (\( p_z > 0 \)). \textit{Inset:} Magnified view of the maximum deflection.  
    \textbf{Right:} Peak shift \( \Delta p_z \) as a function of laser intensity for 800\,nm, 15\,fs pulses. Theoretical results from the Jacobi-Anger (JA) and saddle-point (SP) methods are compared with experimental data \cite{smeenk2011} and prior theoretical predictions \cite{Boning2019}.
	}
\end{figure*}

The figure \ref{Fig_NDshift} provides the evidence of nondipole effects in strong-field ionization, manifested by the forward shift of the photoelectron momentum. In the left panel, both the 800~nm and 1200~nm pulses produce momentum distributions that are asymmetrically shifted toward positive $p_z$, with the longer wavelength (1200~nm) showing a marginally greater maximum deflection.

The right panel comparison of peak shift $\Delta p_z$ across laser intensities further emphasizes the significance of nondipole effects. The Jacobi-Anger (JA) method reproduces the experimental data across the entire intensity range, accurately capturing both the magnitude and increasing trend of $\Delta p_z$. This agreement confirms the JA method's ability to model photon momentum transfer through its Bessel-function formulation, which intrinsically incorporates wavelength scaling without relying on trajectory-level approximations.

In contrast, while the saddle-point method predicts a forward shift, it systematically underestimates $\Delta p_z$ at lower intensities. This discrepancy reveals the limitations of treating magnetic field effects as perturbative corrections to classical trajectories, particularly when the electron's quiver radius becomes comparable to the laser wavelength.

\begin{figure*}
	\centering
	\includegraphics[width=1.0\textwidth]{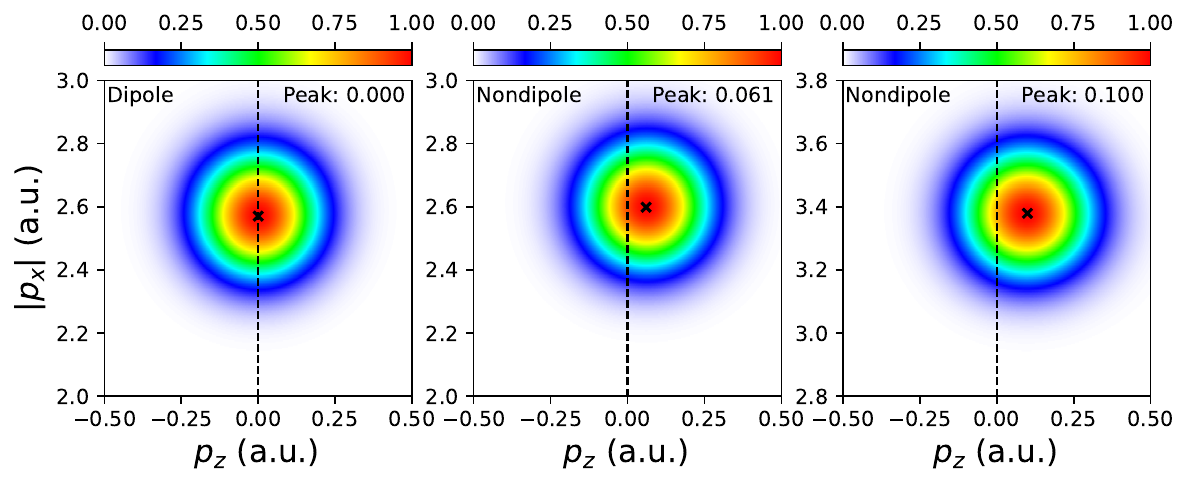}
	\caption{\label{Fig_NDPMD}Comparison of photoelectron momentum distributions in the dipole and nondipole approximations for strong-field ionization of an argon atom with a two-cycle circularly polarized pulse. The distributions are plotted in the \((|p_x|, p_z)\) plane for different laser wavelengths and keeping the intensity constant at \( 5 \times 10^{14} \) W/cm\(^2\). The left panel corresponds to the dipole approximation at \( \lambda = 3200 \) nm, while the middle and right panels show results within the nondipole approximation for \( \lambda = 3200 \) nm and \( \lambda = 4200 \) nm, respectively.  
		The black dashed line marks \( p_z = 0 \), and the peak shift in \( p_z \) due to nondipole effects is indicated in each panel. The black cross denotes the maximum probability location. The observed shift in the peak position along the laser propagation direction (\( p_z \)) highlights the influence of nondipole corrections, which become more pronounced at longer wavelengths. The color scale represents the normalized probability amplitude.  
	}
\end{figure*}

Figure~\ref{Fig_NDPMD} presents a critical analysis of nondipole effects in the strong-field ionization of an argon atom exposed to a two-cycle circularly polarized laser pulse at varying wavelengths. The figure shows the PMDs computed under both the dipole and nondipole approximations, emphasizing the substantial deviations induced by nondipole contributions as the laser wavelength increases. This comparison underscores the importance of accounting for nondipole effects in theoretical and experimental studies of strong-field ionization, particularly in the mid-infrared regime.

In the conventional dipole approximation, depicted in the left panel of Fig.~\ref{Fig_NDPMD} for a wavelength of 3200~nm, the photoelectron momentum distribution exhibits symmetry about the center, $p_z = 0$. This symmetry arises from the assumption that the magnetic field component of the laser pulse has negligible influence on the ionization dynamics. Such an assumption is valid only under specific conditions, such as relatively low laser intensities or shorter wavelengths, where the electric field dominates the interaction. The dipole approximation simplifies the theoretical treatment by neglecting the momentum transfer associated with the magnetic field, thereby reducing computational complexity. However, this simplification becomes increasingly inadequate as the wavelength of the laser pulse increases, necessitating a more comprehensive treatment that includes nondipole effects.
When nondipole effects are incorporated, as illustrated in the middle and right panels of Fig.~\ref{Fig_NDPMD} for wavelengths of 3200~nm and 4200~nm, respectively, a pronounced shift in the peak of the electron momentum distribution is observed. This shift, which is directed towards positive $p_z$ values, becomes more prominent at longer wavelengths, with the peak momentum reaching $p_z = 0.1$~a.u. at 4200~nm. This phenomenon can be attributed to the momentum transfer imparted by the magnetic field of the laser pulse, an effect that is entirely neglected in the dipole approximation. The longitudinal momentum transfer, which arises from the interaction of the photoelectron with the magnetic field component of the laser, becomes increasingly significant at longer wavelengths and higher intensities. This is due to the enhanced role of the magnetic field in influencing the electron dynamics as the laser wavelength extends into the mid-infrared range.

The progressive deviation of the peak momentum from zero in the presence of nondipole effects highlights the additional force exerted on the photoelectrons in the direction of laser propagation. This force, often referred to as the radiation pressure becomes non-negligible under conditions of high laser intensity and long wavelengths. The observed shift in the PMD peak positions serves as a clear indicator of the growing importance of nondipole effects in strong-field ionization processes. This finding is particularly relevant for experiments conducted in the mid-infrared regime, where the influence of the magnetic field cannot be disregarded without introducing significant errors in the interpretation of results.

\section{Conclusion}

In this work, we have theoretically investigated strong-field ionization processes by analyzing PMD and ATI spectra using the strong-field approximation. The study systematically incorporates both dipole and non-dipole effects, employing the saddle point method to extract critical insights into electron dynamics in intense laser fields.

Our results demonstrate that the PMD structure is significantly influenced by the laser pulse parameters, including intensity, wavelength, and ellipticity. The phase structure imprinted on the electron momentum distribution is directly linked to the vector potential of the laser field, revealing strong signatures of field-driven electron dynamics. The role of non-dipole effects becomes prominent in the long-wavelength regime, where photoelectron momentum distributions exhibit shifts along the laser propagation direction due to the breakdown of the dipole approximation.

The ATI spectra reveal intricate interference patterns arising from quantum trajectories of ionized electrons. The interference structures are highly dependent on the laser pulse characteristics, with clear evidence of phase-dependent enhancements and suppression in the energy spectrum. Our findings indicate that nonlinear interference effects play a crucial role in shaping the observed ATI peaks, with distinct signatures that can be used to probe the underlying ionization mechanisms.

Non-dipole corrections introduce measurable deviations in both PMD and ATI spectra, particularly in the high-intensity and long-wavelength regimes. The Lorentz force contribution to the electron motion results in asymmetries that cannot be captured within the dipole approximation, emphasizing the necessity of incorporating non-dipole terms in theoretical models. These corrections are particularly relevant for experimental setups involving mid-infrared or high-intensity laser pulses, where relativistic effects become non-negligible.

A detailed comparison with previous theoretical and experimental studies shows qualitative and quantitative agreement, reinforcing the validity of our approach. However, certain discrepancies, particularly in the treatment of higher-order terms and the role of Coulomb effects post-ionization, highlight the need for further refinement of the theoretical framework. Future research could extend the analysis to include Coulomb corrections and multi-electron interactions, providing a more complete picture of strong-field ionization dynamics.

In conclusion, this study provides a comprehensive theoretical perspective on strong-field ionization beyond the dipole approximation, offering insights that are relevant for both fundamental research and experimental applications. The findings presented here contribute to the ongoing efforts to understand laser-matter interactions in extreme conditions, paving the way for more refined theoretical models and precision measurements in strong-field physics.

\section{Appendix}\label{Appendix}
The arguments of the Bessel function in the transition amplitude \ref{Eq:TransitionAmplitudeJacobi} include terms involving the squared vector potential amplitudes, their respective frequencies, and interaction terms between different frequency components, adjusted by the ellipticity factor. These terms are explicitly given as
\begin{eqnarray}
	x^D &=& \left( \frac{1 - \epsilon^2}{1 + \epsilon^2}\frac{\mathcal{A}_0^2}{8\omega_0}, \frac{1 - \epsilon^2}{1 + \epsilon^2}\frac{\mathcal{A}_1^2}{8\omega_1}, \frac{1 - \epsilon^2}{1 + \epsilon^2}\frac{\mathcal{A}_2^2}{8\omega_2},  \frac{\mathcal{A}_0 \mathcal{A}_1}{2(\omega_0 - \omega_1)}, \frac{\mathcal{A}_0 \mathcal{A}_2}{2(\omega_0 - \omega_2)}, \frac{\mathcal{A}_1 \mathcal{A}_2}{2(\omega_1 - \omega_2)}, \right.  \nonumber \\
	&& \left. \frac{1 - \epsilon^2}{1 + \epsilon^2}\frac{\mathcal{A}_0 \mathcal{A}_1}{2(\omega_0 + \omega_1)}, \frac{1 - \epsilon^2}{1 + \epsilon^2}\frac{\mathcal{A}_0 \mathcal{A}_2}{2(\omega_0 + \omega_2)}, \frac{1 - \epsilon^2}{1 + \epsilon^2}\frac{\mathcal{A}_1 \mathcal{A}_2}{2(\omega_1 + \omega_2)},\quad \right.  \nonumber \\
	&& \left. \frac{p_x}{\sqrt{1 + \epsilon^2}}\frac{\mathcal{A}_0}{\omega_0}, \frac{p_x}{\sqrt{1 + \epsilon^2}}\frac{\mathcal{A}_1}{\omega_1}, \frac{p_x}{\sqrt{1 + \epsilon^2}}\frac{\mathcal{A}_2}{\omega_2},\right.  \nonumber \\
	&& \left. \epsilon \Lambda \frac{p_y}{\sqrt{1 + \epsilon^2}}\frac{\mathcal{A}_0}{\omega_0}, \epsilon \Lambda \frac{p_y}{\sqrt{1 + \epsilon^2}}\frac{\mathcal{A}_1}{\omega_1}, \epsilon \Lambda \frac{p_y}{\sqrt{1 + \epsilon^2}}\frac{\mathcal{A}_2}{\omega_2}, \right),
\end{eqnarray}
where the first set of terms represents the contributions from individual frequency components, while the remaining terms describe cross-interactions between them.
The modified frequency and phase, essential for describing the transition dynamics, are defined as
\begin{eqnarray}
	\Omega^D &=& \left(
	2n_1\omega_0 + 2n_2\omega_1 + 2n_3\omega_2 + n_4(\omega_0 - \omega_1) + n_5(\omega_0 - \omega_2) + n_6(\omega_1 - \omega_2) \right.\nonumber \\
	& & \left. + n_7(\omega_0 + \omega_1) + n_8(\omega_0 + \omega_2) + n_9(\omega_1 + \omega_2) + n_{10}\omega_0 + n_{11}\omega_1 + n_{12}\omega_2 \right.\nonumber \\
	& & \left. + n_{13}\omega_0 + n_{14}\omega_1 + n_{15}\omega_2 \right) \\
	\Phi^D &=& \left(
	2n_1\phi_{\mathrm{cep}} + 2n_2\phi_{\mathrm{cep}} + 2n_3\phi_{\mathrm{cep}} + 2n_7\phi_{\mathrm{cep}} + 2n_8\phi_{\mathrm{cep}} + 2n_9\phi_{\mathrm{cep}} + n_{10}\phi_{\mathrm{cep}} \right.\nonumber \\
	& & \left. + n_{11}\phi_{\mathrm{cep}} + n_{12}\phi_{\mathrm{cep}} + n_{13}\phi_{\mathrm{cep}} + n_{14}\phi_{\mathrm{cep}} + n_{15}\phi_{\mathrm{cep}} \right).
\end{eqnarray}
These expressions account for the summation and combination of different frequency components and their respective phase contributions.

For the nondipole transition amplitude \ref{Eq:TransitionAmplitudeNondipole}, the arguments of the Bessel function take a similar form, incorporating the modified frequencies \(\eta_i(\mathbf{k})\) instead of \(\omega_i\)
\begin{eqnarray}
	x^{ND} &=& \left( 
	\frac{1 - \epsilon^2}{1 + \epsilon^2} \frac{\mathcal{A}_0^2}{8\eta_0(\mathbf{k})}, 
	\frac{1 - \epsilon^2}{1 + \epsilon^2} \frac{\mathcal{A}_1^2}{8\eta_1(\mathbf{k})},
	\frac{1 - \epsilon^2}{1 + \epsilon^2} \frac{\mathcal{A}_2^2}{8\eta_2(\mathbf{k})}, \right. \nonumber \\
	&& \left. \frac{\mathcal{A}_0 \mathcal{A}_1}{2(\eta_0(\mathbf{k}) - \eta_1(\mathbf{k}))},
	\frac{\mathcal{A}_0 \mathcal{A}_2}{2(\eta_0(\mathbf{k}) - \eta_2(\mathbf{k}))},
	\frac{\mathcal{A}_1 \mathcal{A}_2}{2(\eta_1(\mathbf{k}) - \eta_2(\mathbf{k}))}, \right. \nonumber \\
	&& \left. \frac{1 - \epsilon^2}{1 + \epsilon^2} \frac{\mathcal{A}_0 \mathcal{A}_1}{2(\eta_0(\mathbf{k}) + \eta_1(\mathbf{k}))},
	\frac{1 - \epsilon^2}{1 + \epsilon^2} \frac{\mathcal{A}_0 \mathcal{A}_2}{2(\eta_0(\mathbf{k}) + \eta_2(\mathbf{k}))},
	\frac{1 - \epsilon^2}{1 + \epsilon^2} \frac{\mathcal{A}_1 \mathcal{A}_2}{2(\eta_1(\mathbf{k}) + \eta_2(\mathbf{k}))}, \right. \nonumber \\
	&& \left. \frac{p_x}{\sqrt{1 + \epsilon^2}} \frac{\mathcal{A}_0}{\eta_0(\mathbf{k})},
	\frac{p_x}{\sqrt{1 + \epsilon^2}} \frac{\mathcal{A}_1}{\eta_1(\mathbf{k})},
	\frac{p_x}{\sqrt{1 + \epsilon^2}} \frac{\mathcal{A}_2}{\eta_2(\mathbf{k})}, \right. \nonumber \\
	&& \left. \epsilon \Lambda \frac{p_y}{\sqrt{1 + \epsilon^2}} \frac{\mathcal{A}_0}{\eta_0(\mathbf{k})},
	\epsilon \Lambda \frac{p_y}{\sqrt{1 + \epsilon^2}} \frac{\mathcal{A}_1}{\eta_1(\mathbf{k})},
	\epsilon \Lambda \frac{p_y}{\sqrt{1 + \epsilon^2}} \frac{\mathcal{A}_2}{\eta_2(\mathbf{k})} \right).
\end{eqnarray}
The quantities \(\eta_{i}(\mathbf{k})\) for \(i = 0, 1, 2\) are defined as \(\eta_{i}(\mathbf{k}) = \mathbf{p} \cdot \mathbf{k}_i - \omega_i\), where \(\mathbf{p}\) is the momentum vector, and \(\mathbf{k}_i\) and \(\omega_i\) are the wave vector and angular frequency associated with the \(i\)-th component, respectively. Specifically, these relations are expressed as \(\eta_{i}(\mathbf{k}) = \left(\frac{p_z}{c} - 1\right)\omega_i\), with \(p_z\) being the component of \(\mathbf{p}\) along the \(z\)-axis, and \(c\) representing the speed of light in vacuum. Finally, the modified frequency, phase and momentum in the nondipole case are given by
\begin{eqnarray}
	\Omega^{ND} &=& -\left(
	2n_1\omega_0 + 2n_2\omega_1 + 2n_3\omega_2 + n_4(\omega_0 - \omega_1) + n_5(\omega_0 - \omega_2) + n_6(\omega_1 - \omega_2) \right.\nonumber \\
	& & \left. + n_7(\omega_0 + \omega_1) + n_8(\omega_0 + \omega_2) + n_9(\omega_1 + \omega_2) + n_{10}\omega_0 + n_{11}\omega_1 + n_{12}\omega_2 \right.\nonumber \\
	& & \left. + n_{13}\omega_0 + n_{14}\omega_1 + n_{15}\omega_2 \right) \\
	\Phi^{ND} &=& \left(
	2n_1\phi_{\mathrm{cep}} + 2n_2\phi_{\mathrm{cep}} + 2n_3\phi_{\mathrm{cep}} + 2n_7\phi_{\mathrm{cep}} + 2n_8\phi_{\mathrm{cep}} + 2n_9\phi_{\mathrm{cep}} + n_{10}\phi_{\mathrm{cep}} \right.\nonumber \\
	& & \left. + n_{11}\phi_{\mathrm{cep}} + n_{12}\phi_{\mathrm{cep}} + n_{13}\phi_{\mathrm{cep}} + n_{14}\phi_{\mathrm{cep}} + n_{15}\phi_{\mathrm{cep}} \right)\\
	\mathbf{p}_{n_i} &=& \mathbf{p} + \left(\sum_{j=0}^{2}\frac{\mathcal{A}_j}{4\eta_j} - \frac{\Omega^{ND}}{\omega}\right)\mathbf{k} 
\end{eqnarray}
\section{References}
\bibliographystyle{iopart-num}
\bibliography{sfa}

\end{document}